\documentclass[10pt,final,a4paper,twocolumn,twoside,journal,romanappendices]{IEEEtran}
\usepackage{cite}
\usepackage{epstopdf}
\usepackage{epsfig}
\usepackage{amsmath,amssymb,amsfonts,amstext,amsbsy,amsopn,dsfont}
\usepackage{cases}
\usepackage{color}
\usepackage{stfloats}
\usepackage{sublabel}
 \usepackage{bigints}
\usepackage{array}

\newtheorem{theorem}{\textbf{Theorem}}

\newtheorem{remark}{\textbf{Remark}}

\newcommand{\defn}{\triangleq}
\newcommand{\dif}{\textmd{d}}

\begin{document}

\title{ Fundamentals of Simultaneous Wireless Information and Power Transmission in Heterogeneous Networks: A Cell Load Perspective}

\author{Chun-Hung Liu, \IEEEmembership{Senior Member, IEEE} and Chi-Sheng Hsu
	\thanks{C.-H. Liu is with the Department of Electrical and Computer Engineering, Mississippi State University, Mississippi State, MS 39762, USA. (e-mail: chliu@ece.msstate.edu)}
	\thanks{C.-S. Hsu is currently with Realtek Semiconductor Inc., Hsinchu 30010, Taiwan. He was with the Institute of Communications Engineering and Department of Electrical and Computer Engineering, National Chiao Tung University, Hsinchu 30010, Taiwan. (e-mail: jacky0560249@gmail.com)}}	


\maketitle

\begin{abstract}
In a heterogeneous cellular network (HetNet) consisting of multiple different types (tiers) of base stations (BSs), the void cell event in which a BS does not have any users has been shown to exist due to user-centric BS association and its probability is dominated by the cell load of each tier. Such a void cell phenomenon has not been well characterized in the modeling and analytical framework of simultaneous wireless information and power transmission (SWIPT) in a HetNet. This paper aims to accurately exploit the fundamental performance limits of the SWIPT between a BS and its user by modeling the cell-load impact on the downlink and uplink transmissions of each BS. We first characterize the power-splitting receiver architecture at a user and analyze the statistical properties and limits of its harvested power and energy, which reveals how much of the average energy can be harvested by users and how likely the self-powered sustainability of users can be achieved. We then derive the downlink and uplink rates that characterize the cell-load and user association effects and use them to define the energy efficiency of a user. The optimality of the energy efficiency is investigated, which maximizes the SWIPT performance of the receiver architecture for different user association and network deployment scenarios. 
\end{abstract}


\section{Introduction}
\IEEEPARstart{S}{imultaneous wireless} information and power transmission (SWIPT) refers to the scenario whereby a transmitter emits electromagnetic waves to its intended receiver that can exploit the power of the electromagnetic waves for information decoding and energy harvesting at the same time. SWIPT is a fairly promising transmission technique for cellular networks because wireless users with a capacity-limited battery can receive information and replenish their battery energy without using external power sources at the same time  \cite{XLPWDNDKZH15}. Cellular networks are gradually migrating to have an architecture with heterogeneity and densification in order to support the explosive traffic demand anticipated in the near future. Since a heterogeneous cellular network (HetNet) can have a considerable wireless power-loss reduction, the HetNet with the SWIPT technology certainly brings about a new wireless communication era in which wireless mobile devices can harness energy from their ambient strong radio frequency (RF) signals emitted by BSs and they could thus no longer operate under the threat of power outage \cite{TDPDKJSKSSCJL18}. Nonetheless, there exist some technical problems that hinder to effectively implement SWIPT in a HetNet, such as energy-harvesting transceiver and algorithm design, energy-efficient network deployment for SWIPT, system integration and protocol design \cite{DNDIKMMZH17}, etc. 

Since there is a minimum power requirement to activate the energy harvesting circuit in a receiver, it is necessary to deploy enough BSs so as to make the receiver acquire sufficiently large electromagnetic  power which can be harvested for some energy. Accordingly, the most challenging problem from the system-level point of view is how to properly deploy base stations in a HetNet so as to efficiently improve the SWIPT performance as much as possible. To tackle this problem, we first need to adopt a general and appropriate model that can characterize the features of a HetNet with SWIPT so that the performance metrics regarding SWIPT can be well defined and tractably analyzed. Charactering the analytical results of the performance metrics for SWIPT is very important in that they not only reveal the fundamental limits on the performance of a HetNet with SWIPT but also shed light on how to achieve the limits by optimally deploying and operating the HetNet.

In the past few years, we have witnessed a great success in modeling a HetNet and analyzing its performance metrics by using stochastic geometry \cite{HSDRKGFBJGA12,PXCHLJGA13,CHLKLF16,CHLHCT17,HEASSMS17}. As such, to generally and tractably delve the fundamental performances of a HetNet with SWIPT, in this paper we still adopt the stochastic geometry framework to model $M$ different types of BSs equipped multiple antennas in a large-scale HetNet as $M$ independent Poisson point processes (PPPs). Each user in the HetNet is assumed to be equipped with a single antenna. Namely, our current goal is to tractably exploit the fundamental performances of a multiple-input-single-output (MISO) channel for SWIPT\footnote{Due to the intractability in analyzing random matrices in a HetNet, in this paper our study focuses on a MISO channel for SWIPT  instead of a MIMO channel.}. For a receiver with a single antenna, principally it has two types of architecture for SWIPT, i.e., time-switching architecture and power-splitting architecture \cite{TDPDKJSKSSCJL18}\cite{YASRCYL18}. Since the power-splitting architecture theoretically achieves the best trade-off performance between information decoding and energy harvesting \cite{XZRZCKH13}, in this paper we adopt it in a receiver and create the received power models for information decoding and energy harvesting, respectively.

\subsection{Prior Work and Motivation}

Wireless powered communication networks (WPCNs) have gained a lot of attentions in recent years since BSs (or power beacons) in the networks are convenient wireless power sources to replenish the power consumption of users. Most of the prior works on WPCNs were accomplished merely based on a single-cell network model\cite{KHEL13,HJRZ14,TRAJ17}. In \cite{HJRZ14} and \cite{TRAJ17}, for example, the uplink throughput maximization problem with wireless power transfer constraints in the downlink was studied for a single-cell network. Since such a single-cell network model ignores the interferences from other cells, the harvested energy could be largely underestimated because the interferences from other nearby BSs could be strong energy sources to be harvested, especially when the entire network is densely deployed. As such, a dense HetNet is doubtlessly an excellent environment for wireless power transmission so that wireless powered communication in HetNets has recently become an active research topic and some typical prior works on this topic can be referred to \cite{YDLWMEMD16,YZLWKWSJ16,JHPYSJSH17,SHKDIK17}. These prior works mainly aimed to investigate how much averaged power can be harvested at users and how to  optimally use multiple antennas to effectively deliver power to users. Although they indeed considered a more general multi-cell network model, there are still some important issues that were circumvented by their modeling assumptions, such as how different user association schemes affect their analyses and how user scheduling impacts the efficiency of wireless powered communications, etc. 

In addition to WPCNs, the aforementioned SWIPT is also a very promising technique for future wireless networks. This is attributed to its potential capability of effectively transmitting information and power at the same time if the trade-off between information decoding and energy harvesting can be cleverly overcome. Some prior works on SWIPT in large-scale wireless networks (typically see \cite{IK14,SLEHVKB15,SAYDANME16,MSLWXWYZ16,MDRWL17,HZJDJCKLVL18}) had already studied the popular performance metrics, such as coverage/outage probability, downlink rate, average received energy and energy efficiency. For instance, reference \cite{IK14} analyzed the outage probability performance and the averaged harvested energy for a power-splitting receiver in a wireless ad hoc network using non-cooperative and cooperative relaying protocols, whereas it did not study how transmission scheduling and relay selection impact the outage and energy-harvesting performances. In \cite{MDRWL17}, the distributions of the harvested power and downlink rate for different SWIPT techniques were analyzed in a general mathematical approach under a realistic channel model. Reference \cite{SAYDANME16} studied the SWIPT problem in a $K$-tier HetNet and analyzed the downlink outage and rate, whereas the statistical properties of the harvested power were not addressed. Among these prior works, all of their analytical results were obtained by assuming a specific user association scheme is adopted and all BSs are actively working. Such assumptions may be detached from the reality and could lead to inaccurate analysis for a dense HetNet in which BSs may not be always active due to lack of users \cite{CHLLCW1502}\cite{CHLLCW16}. 
 
\subsection{Main Contributions}
There is a crucial issue that is commonly overlooked in the prior works on the modeling and analysis of a HetNet with SWIPT, i.e., the cell load impacts on SWIPT have not been discovered and included in the network model. For an $M$-tier HetNet, the tier-$m$ cell load represents the average number of users associating with a tier-$m$ BS \cite{CHLKLF16}\cite{CHLHCT1702} and it significantly affects how likely a tier-$m$ BS is associated with at least one user according to our previous works in \cite{CHLLCW1502}\cite{CHLLCW16}. This finding discloses the fact that there indeed exist a certain number of the \textit{void} BSs that do not have any users and no energy can be harvested from these void BSs. Our main contributions in this paper are to fundamentally study how the cell loads in different tiers influence the SWIPT performances in terms of the harvested energy, rate and energy efficiency. They are briefly summarized in the following:
 \begin{itemize}
 	\item We model the HetNet with BS voidness and propose a received power model that characterizes the phenomenon of void BSs, and we then find the statistical properties of  a generalized $n$th-incomplete shot noise process and apply them to characterize the received power from a downlink MISO channel. 
 	\item The tight lower bounds on the Laplace transform and the cumulative distribution function (CDF) of the harvested power at a receiver with a power-splitting architecture are found for a generalized user association scheme. They results in the finding of the fundamental lowest limit on the CDF of the harvested power as users associate their strongest BS and all cell loads go to infinity. 
 	\item The outage probability of energy harvesting is studied, which is a good index to indicate how likely the received power is too weak to be harvested at a receiver. Moreover, we study the outage probability of self-powered sustainability that indicates how likely the energy harvested from SWIPT is large enough to completely support the energy needed for uplink transmission.
 	\item We formulate the (ergodic) downlink rate for a MISO channel with SWIPT and the uplink rate for a single-input-multiple-output (SIMO) channe and neatly derive their tight lower bounds. We thus are able to characterize the fundamental limits on the link rates when the cell loads as well as the number of antennas of BSs go to infinity. 
 	\item We define the energy efficiency as the sum rate in the downlink and uplink divided by the total energy consumed in the downlink and uplink and formulate an optimization problem to maximize it with the constraints on the outage probability of energy harvesting and self-powered sustainability. We show that the optimal power-splitting factor and the optimal downlink time fraction for a SWIPT receiver can be analytically found provided the parameters of a HetNet satisfy with the derived constraints. 
 \end{itemize}
\noindent Furthermore, some numerical simulation results are provided to validate our analytical findings and observations.

 \subsection{Paper Organization}
 The rest of this paper is organized as follows. In Section \ref{Sec:SystemModel}, the system model and assumptions for a HetNet with SWIPT are elaborated. Section \ref{Sec:AnalysisHarvestedPower} analyzes the statistical properties of the harvested power and energy and it also provides some numerical results to validate the analytical findings. In Section \ref{Sec:LinkRateEnergyEff}, the downlink and uplink rates for SWIPT are first defined and analyzed. Afterwards, the energy efficiency of a user that is defined based on the link rates is analyzed and its optimality is investigated. Finally, Section \ref{Sec:Conclusion} concludes our analytical observations and findings.

\section{System Model and Assumptions}\label{Sec:SystemModel}
In this paper, we consider a large-scale HetNet on the $\mathbb{R}^2$ plane in which there are $M$ different types of BSs, e.g., macrocell, picocell and femtocell BSs. The BSs of each specific type are referred to as a tier of the HetNet and they form an independent and homogeneous Poisson point process (PPP). Specifically, the BSs in the $m$th tier have intensity $\lambda_m$ and they can be expressed as set $\mathcal{B}_m$ given by
\begin{align}
\mathcal{B}_m\defn\left\{B_{m,i}\in\mathbb{R}^2: j\in\mathbb{N}\right\},
\end{align}
where $m\in\mathcal{M}\defn\{1,2,\ldots,M\}$, $B_{m,i}$ denotes BS $i$ in the $m$th tier and its location in the HetNet. Each tier-$m$ BS has transmit power $P_m$ and is equipped with $N_m$ transmit antennas. Full-frequency reuse is adopted in the HetNet so as to enhance the spectrum efficiency. Also, all users form an independent homogeneous PPP of intensity $\mu$, which is denoted by set $\mathcal{U}$ in the following:
\begin{align}
\mathcal{U}\defn\{U_j\in\mathbb{R}^2: j\in\mathbb{N}\}.
\end{align}
All users have the same transmit power $Q$ and are equipped with a single antenna\footnote{ To make the analyses much tractable in this paper, users are only considered to be equipped with a single antenna so that all analyses in the downlink are performed based on the MISO channel model, whereas all analyses in the uplink are proceeded based on the SIMO channel model. Nevertheless, all the analytical results are scalable to their corresponding counterparts with miltiple-input-multiple-output (MIMO) channels by using proper scaling techniques. }. 

\subsection{User Association Scheme and Its Related Statistics}
Each user in set $\mathcal{U}$ selects its serving BS by adopting the following \textit{generalized user association (GUA) scheme}
\begin{align}\label{Eqn:UserAssScheme}
B_*\defn\arg\max_{m,i:B_{m,i}\in\mathcal{B}}\left\{w_m|B_{m,i}|^{-\alpha}\right\},
\end{align}
where $B_*$ denotes the BS associated with a user located at the origin, $\mathcal{B}\defn\bigcup_{m=1}^M \mathcal{B}_m$ is the set of all BSs in the HetNet, $w_m>0$ is called the tier-$m$ user association weight, $|X-Y|$ denotes the Euclidean distance between nodes $X$ and $Y$, and $\alpha>2$ is the path-loss exponent\footnote{In practice, the path-loss exponents of the path-loss models in different tiers should differ from one another. However, in this paper we still decide to adopt the same path-loss exponent for different tiers since using different path-loss exponents leads to high intractability in the following analyses and we thus cannot derive any insightfully analytical results.}. Note that for simplicity we use the location of the origin to specify the user association scheme in \eqref{Eqn:UserAssScheme} since the Slinvyak theorem indicates that  the statistical properties of a homogeneous PPP evaluated at any particular point in the network are the same as those evaluated at any other locations in the same network \cite{DSWKJM13}. The scheme in \eqref{Eqn:UserAssScheme} is so general that it is able to cover several user association schemes \cite{CHLLCW16,CHLKLF16}. For example, users associate with their nearest BS if all $w_m$'s are equal to unity, which is called the \textit{nearest BS association} (NBA) scheme. When users associate with their strongest BS by averaging out channel fading variations (i.e., $w_m\equiv P_m$ for all $m\in\mathcal{M}$), this scheme is called the \textit{maximum (mean) received power association} (MRPA) scheme. According to Lemma 1 in our previous work \cite{CHLKLF16}, the probability mass function of the number of users associating with a tier-$m$ BS for the user association scheme in \eqref{Eqn:UserAssScheme} can be inferred as
\begin{align}\label{Eqn:ProbMassFunNumUser}
p_{m,n}=\frac{\Gamma(n+\frac{7}{2})}{n!\Gamma(\frac{7}{2})}\left(\frac{2}{7}\ell_m\right)^n\left(1+\frac{2}{7}\ell_m\right)^{-(n+\frac{7}{2})},
\end{align}
where $\Gamma(a)\defn\int_{0}^{\infty} t^{a-1}e^{-t}\dif t$ is the Gamma function, $\ell_m\defn w^{\frac{2}{\alpha}}_m\mu/\lambda_{\Sigma}$
is the tier-$m$ \textit{cell load} that indicates the average number of users associating with a tier-$m$ BS, and $\lambda_{\Sigma}\defn \sum_{m=1}^{M}w^{\frac{2}{\alpha}}_m\lambda_m$ is called the \textit{equivalent sum intensity of BSs for the GUA scheme} in \eqref{Eqn:UserAssScheme}. The result in \eqref{Eqn:ProbMassFunNumUser} reveals that the tier-$m$ non-void probability that there is at least one user associating with a tier-$m$ BS is given by
\begin{align}
q_m\defn 1-p_{m,\emptyset} =1-\left(1+\frac{2}{7}\ell_m\right)^{-\frac{7}{2}},
\end{align}
which is small whenever $\ell_m$ is large. In other words, there are a large number of void tier-$m$ BSs in the HetNet whenever the tier-$m$ cell load is small. Thus, we need to be aware of how the void BSs that do generate interference influence the performance metrics while doing modeling and analysis in the following. In addition, let $w_*|B_*|^{-\alpha}\defn \max_{B_{m,i}\in\mathcal{B}}\left\{w_m|B_{m,i}|^{-\alpha}\right\}$ where $w_*\in\{w_1,\ldots,w_M\}$ denotes the user association weight used by $B_*$. The CDF of $w^{-\frac{2}{\alpha}}_*|B_*|^2$, based on the result in \cite{CHLKLF16}, can be found as
\begin{align}\label{Eqn:CDFAssBS}
F_{w^{-\frac{2}{\alpha}}_*|B_*|^2}(\theta) &\defn\mathbb{P}\left[w^{-\frac{2}{\alpha}}_*|B_*|^2\leq \theta\right]\nonumber\\
&=1-\exp\left(-\pi \lambda_{\Sigma}\theta\right),
\end{align}
where $F_Z(\cdot)$ denotes the CDF of random variable (RV) $Z$. Also, the time division duplex (TDD) mode is adopted in the HetNet and time division multiple access (TDMA) is used to schedule the transmission of users so that each user is able to utilize all resources from its serving BS within its scheduled time slot and there thus is no intra-cell interference. Moreover, all scheduled users are assumed to alway have data to send to their BS in the uplink  time slot. 

\subsection{Model of Simultaneous Wireless Information and Power Transmission (SWIPT) }\label{SubSec:SWIPT}

Suppose the BS of the typical user located at the origin can send information to it by using transmit beamforming. The model of the total received signal power for the user can be expressed as follows\footnote{To simplify the notation in \eqref{Eqn:ReceivedDownLinkPower}, the received power $P_{dl}$ is normalized by the path-loss coefficient at a reference distance of 1 meter.}:
\begin{align}\label{Eqn:ReceivedDownLinkPower}
P_{dl}\defn P_*H_*\|B_*\|^{-\alpha}+I_{dl},
\end{align}
where  $\|X-Y\|^{-\alpha}\defn |X-Y|^{-\alpha}\mathds{1}(|X-Y|\geq 1)$ for all $X, Y\in\mathbb{R}^2$, $\mathds{1}(\mathcal{E})$ is the indicator function that is equal to unity if event $\mathcal{E}$ is true and zero otherwise,  the term $P_*H_*\|B_*\|^{-\alpha}$ is the desired signal power\footnote{The motivation of using the path-loss model $\|\cdot\|^{-\alpha}$ is due to the fact that the model $|\cdot|^{-\alpha}$ does not behave well in the near field of a transmitter and it thus leads to an unbounded mean of the shot noise process such as $I_{dl}$. This path-loss model follows the idea of the bounded propagation model proposed in \cite{CHLJGA12} and it is still an accurate model for doing analysis in a Poisson network since the node intensity in such a network is usually fairly small.}, $P_*\in\{P_1,\ldots,P_M\}$ is the transmit power of BS $B_*$, $H_*$ is the channel fading gain between the typical user and BS $B_*$, and $I_{dl}$ is the interference power given by
\begin{align}\label{Eqn:DownlinkInterference}
I_{dl} \defn \sum_{m,i:B_{m,i}\in\mathcal{B}\setminus B_*} P_mV_{m,i}H_{m,i}\|B_{m,i}\|^{-\alpha}
\end{align}
in which $V_{m,i}\in\{0,1\}$ is a Bernoulli random variable (RV) that is unity if BS $B_{m,i}$ is not void and zero otherwise. Note that the received thermal noise is ignored in \eqref{Eqn:ReceivedDownLinkPower} since it is usually very much smaller than  the received interference power. All channel gains between users and their BSs undergo identical and independent Rayleigh fading and we assume that $H_{m,i}\sim\exp(1)$ is an exponential RV with unit mean and variance for all $m\in\mathcal{M}$ and $i\in\mathbb{N}_+$. All channel gains are also block-fading, i.e., they are independent in different time slots.  

\begin{figure}[!t]
	\centering
	\includegraphics[width=3.3in,height=2.3in]{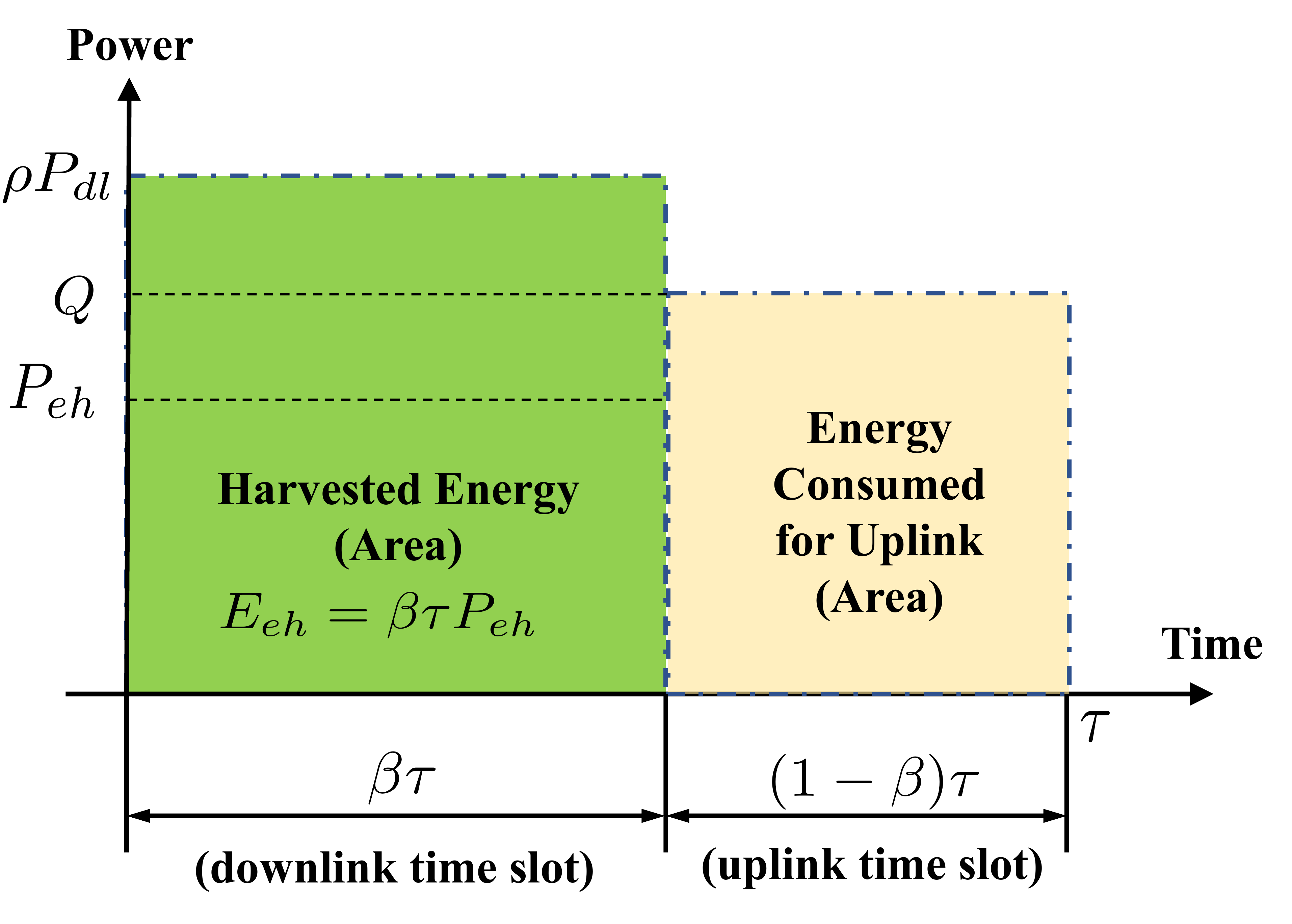}
	\caption{An illustration of the SWIPT model used in this paper. In the figure, $\tau$ is the total time duration of the downlink and uplink transmissions of a users in the HetNet, $\beta\in(0,1)$ is the time fraction for downlink transmission, $P_{dl}$ denotes the total received power at a user, $\rho P_{dl}$ denotes the received power with power splitting factor $\rho\in(0,1)$ for information decoding,  $P_{eh}=\eta(1-\rho)P_{dl}$ is the converted harvested power from the energy harvesting circuits with efficiency $\eta\in(0,1)$, $Q$ is the transmit power of users, and $E_{eh}$ is the harvested energy.}
	\label{Fig:SwiptModel}
\end{figure}

The SWIPT model between a user and its tagged BS is specified as follows. Suppose the total transmission duration for downlink and uplink is $\tau\in\mathbb{R}_+$. Let $\beta\in(0,1)$ be the time fraction for downlink, which means that $\beta \tau$ and $(1-\beta)\tau$ are the downlink transmission time for each BS and the uplink transmission time for each user, respectively. Each user has a battery with large capacity which can storage the energy harvested from the received RF power signals. The harvest-then-transmit protocol is used in the HetNet, that is, users are able to harvest the transmitted power from its tagged BS during the downlink transmission time period of $\beta\tau$ and then transmit its data to its BS during the uplink transmission time period of $(1-\beta)\tau$. Furthermore, the receiver of each user is assumed to have a power-splitting architecture with a power splitting factor $\rho\in(0,1)$ which is able to split the total received power $P_{dl}$ into two powers $\rho P_{dl}$ and $(1-\rho)P_{dl}$: The power $\rho  P_{dl}$ is for information decoding, whereas the power $(1-\rho)P_{dl}$ is for energy harvesting. An illustration of this SWIPT model is depicted in Fig. \ref{Fig:SwiptModel}. This SWIPT induces a special signal power structure whose statistical properties can be characterized by the $n$th generalized incomplete shot noise process introduced in the following subsection. 

\subsection{Analysis of the Generalized $n$th-Incomplete Shot Noise Process}
Consider a homogeneous PPP of intensity $\lambda_{\mathcal{X}}$ that is denoted by set $\mathcal{X}\defn\{X_n\in\mathbb{R}^2: n\in\mathbb{N}_+\}$ in which $X_n$ is the $n$th nearest point in $\mathcal{X}$ to the origin. For set $\mathcal{X}$, its generalized  $n$th-incomplete shot noise process is defined as
\begin{align}\label{Eqn:GenIncomShotNoise}
I_{(n)}\defn \widehat{W}_n\|X_n\|^{-\alpha}+\sum_{n:X_{n+1}\in\mathcal{X}}W_{n+1}\|X_{n+1}\|^{-\alpha},
\end{align}
where $\widehat{W}_n$ is a nonnegative RV with certain distribution and all $W_{n+1}$'s are i.i.d. nonnegative RVs for all $n\in\mathbb{N}$. Let $\mathcal{L}_Z(s)\defn \mathbb{E}[e^{-sZ}]$ for $s>0$ be the Laplace transform of a nonnegative RV $Z$ and thereby we have the following theorem.
\begin{theorem}\label{Thm:LapMeanNthShotNoise}
If the Laplace transforms and means of $\widehat{W}_n$ and $W_n$ in \eqref{Eqn:GenIncomShotNoise} all exist, the Laplace transform of the $n$th generalized incomplete shot noise process $I_{(n)}$ in \eqref{Eqn:GenIncomShotNoise} is explicitly found as
\begin{align}\label{Eqn:LapGenNthShotNoise}
\mathcal{L}_{I_{(n)}} (s) =& e^{-\pi\lambda_{\mathcal{X}}s^{\frac{2}{\alpha}}\Gamma\left(1-\frac{2}{\alpha}\right)\mathbb{E}\left[W^{\frac{2}{\alpha}}\right]}\frac{(\pi\lambda_{\mathcal{X}})^n}{(n-1)!}\times\nonumber\\
&\int_{1}^{\infty}\mathcal{L}_{\widehat{W}_n}\left(s^{-\frac{\alpha}{2}}x\right)e^{-\pi\lambda_{\mathcal{X}}\int_{0}^{x} \mathcal{L}_W\left(sv^{-\frac{\alpha}{2}}\right)\dif v}x^{n-1}\dif x,
\end{align} 
where $n\in\mathbb{N}_+$. Also, the mean of $I_{(n)}$ can be found as
\begin{align}\label{Eqn:MeanGenNthShotNoise}
\mathbb{E}\left[I_{(n)}\right] 
=&\frac{(\pi\lambda_{\mathcal{X}})^{\frac{\alpha}{2}}}{(n-1)!} \bigg[ \left(\mathbb{E}\left[\widehat{W}_n\right]-\frac{(\alpha-2n)\mathbb{E}\left[W\right]}{(\alpha-2)}\right)\times\nonumber\\
&\Gamma\left(n-\frac{\alpha}{2},\pi\lambda_{\mathcal{X}}\right)+\frac{2\mathbb{E}\left[W\right](\pi\lambda_{\mathcal{X}})^{n-\frac{\alpha}{2}}}{(\alpha-2)e^{\pi\lambda_{\mathcal{X}}}}\bigg],
\end{align}
where $\Gamma(a,b)=\int_{b}^{\infty} t^{a-1}e^{-t}\dif t$ is the upper incomplete Gamma function for $b\geq 0$. 
\end{theorem}
\begin{IEEEproof}
See Appendix \ref{App:ProofLapMeanNthShotNoise}.
\end{IEEEproof}
Although the results in \eqref{Eqn:LapGenNthShotNoise} and \eqref{Eqn:MeanGenNthShotNoise} are somewhat complex, they are very general and they can be largely simplified if we consider some special cases. For example, considering $n=1$, $W_n\sim\exp(1)$ and $\alpha=4$, yields the following results:
\begin{align}
\mathcal{L}_{I_{(1)}} (s) =\pi\lambda_{\mathcal{X}} \sqrt{s}\int_{1}^{\infty} \mathcal{L}_{\widehat{W}_1}\left(y\right)e^{-\pi\lambda_{\mathcal{X}}\sqrt{s}\left[y-\tan^{-1}\left(y\right)+\frac{\pi}{2}\right]} \dif y
\end{align}
and
\begin{align}
\mathbb{E}\left[I_{(1)}\right] 
&=\frac{\pi\lambda_{\mathcal{X}}}{2}e^{-\pi\lambda_{\mathcal{X}}},
\end{align}
which is close to $\frac{\pi}{2}\lambda_{\mathcal{X}}$ if $\lambda_{\mathcal{X}}\ll 1$. As can be seen in the following sections, the general results in Theorem \ref{Thm:LapMeanNthShotNoise} play a pivotal role in facilitating many derivations in the following analyses of the link rates and harvested power and energy.  

\section{Analysis of the Harvested Power and Energy}\label{Sec:AnalysisHarvestedPower}
As mentioned in Section \ref{SubSec:SWIPT}, the received signal power for energy harvesting is $(1-\rho) P_{dl}$. Let $\eta\in(0,1)$ be the energy conversion efficiency of the energy harvesting circuits at users\footnote{In practice, the energy conversion efficiency $\eta$ may not be a constant since it may change over time due to the non-linearity property of the EH circuits at a user while the user is harvesting energy. However, in this paper we consider a constant $\eta $ because our focus is to analyze the harvested power in a time-averaged sense.} and the harvested power can thus be simply expressed as
\begin{align}\label{Eqn:HavestedPower}
P_{eh}\defn \eta (1-\rho) P_{dl}.
\end{align} 
According to \eqref{Eqn:ReceivedDownLinkPower}, $P_{eh}$ is a random variable depending upon user association, channel fading, distribution of BSs. Its distribution significantly influences the performance of energy harvesting at the receiver of users. In the following subsection, the statistical properties of the harvested energy are first investigated and other properties of the harvested energy will then be exploited. 

\subsection{The Statistical Properties of the Harvested Power}
The explicit result regarding the Laplace transform and the CDF of $P_{eh}$ can be found as shown in the following theorem.
\begin{theorem}\label{Thm:CDFHarvEng}
Suppose the fading channel gain $H_*$ is a Gamma RV with shape and scale parameters $N_m$ and $1/N_m$ (i.e., $H_*\sim\text{Gamma}(N_m,1/N_m)$)  whenever $B_*\in\Phi_m$. If the user association scheme in \eqref{Eqn:UserAssScheme} is adopted, then the Laplace transform of the harvested power in \eqref{Eqn:HavestedPower} is tightly lower bounded by
\begin{align}\label{Eqn:LapTransHavEng}
\mathcal{L}_{P_{eh}}(s) \gtrapprox& \sum_{m=1}^{M}\vartheta_m\bigintssss_{0}^{\infty}\pi\lambda_{\Sigma}\left(1+\frac{\eta(1-\rho)sP_m}{w_mN_mx^{\frac{\alpha}{2}}}\right)^{-N_m}\times\nonumber\\
&e^{-\pi \lambda_{\Sigma} \left( \sum_{m=1}^{M}\Phi_m\left(x,s\right)+x\right) }\dif x,
\end{align} 
where $a\gtrapprox b$ means $b$ is a tight lower bound on $a$, function $\Phi_m(y,z)$ is defined as
\begin{align}\label{Eqn:FunPhi}
\Phi_m (y,z)\defn &  \vartheta_mq_m\left(\frac{\eta(1-\rho)zP_m}{w_m}\right)^{\frac{2}{\alpha}}\times\nonumber\\ 
&\left(\frac{2\pi/\alpha}{\sin(2\pi/\alpha)}- \int_{0}^{y\left(\frac{w_m}{\eta(1-\rho)zP_m}\right)^{\frac{2}{\alpha}}} \frac{\dif t}{1+t^{\frac{\alpha}{2}}}\right),
\end{align}
and $\vartheta_m\defn w^{\frac{2}{\alpha}}_m\lambda_m/\lambda_{\Sigma}=\ell_m\lambda_m/\mu$. By using \eqref{Eqn:LapTransHavEng}, an accurate tight lower on the CDF of $P_{eh}$  bound can be found as
\begin{align}\label{Eqn:CDFHarvEng}
F_{P_{eh}}(\theta) \gtrapprox \bigintssss_{0}^{\infty}\pi\lambda_{\Sigma}\left(1+\frac{\eta(1-\rho)P_m}{w_mN_mu^{\frac{\alpha}{2}}}\right)^{-N_m}\Psi(\theta,u)\,\dif u,
\end{align}
where $\Psi(\theta,u)\defn \mathcal{L}^{-1}\left\{s^{\frac{2}{\alpha}-1}e^{-\pi\lambda_{\Sigma}s^{\frac{2}{\alpha}}\left[\sum_{m=1}^{M}\Phi_m\left(u,1\right)+u\right]}\right\}(\theta)$ and $\mathcal{L}^{-1}(\cdot)(\theta)$ denotes the operator of the inverse Laplace transform with parameter $\theta$. In particular, the result in \eqref{Eqn:CDFHarvEng} for $\alpha=4$ reduces to
\begin{align}\label{Eqn:CDFHarEngAlp4}
F_{P_{eh}}(\theta) \gtrapprox& \sum_{m=1}^{M} \frac{2\vartheta_m}{\sqrt{\pi}}\bigintssss_{0}^{\infty} \left(1+\frac{\pi^2\lambda^2_{\Sigma}\eta(1-\rho)P_m}{4w_mN_m\theta v^2}\right)^{-N_m}\times\nonumber\\
& e^{-\left(\frac{\pi\lambda_{\Sigma}}{2\sqrt{\theta}}\sum_{m=1}^{M}\phi_m(v)+v\right)^2}\dif v,
\end{align}
where $\phi_m(v)$ is given by
\begin{align}
\phi_m(v) =& \vartheta_mq_m\sqrt{\frac{\eta(1-\rho)P_m}{w_m}}\times\nonumber\\ &\left[\frac{\pi}{2}- \tan^{-1}\left(\frac{2v}{\pi\lambda_{\Sigma}}\sqrt{\frac{\theta w_m}{\eta(1-\rho)P_m}}\right)\right]
\end{align}
in which $\lambda_{\Sigma}=\sum_{k=1}^{M}\sqrt{w_k}\lambda_k$.
\end{theorem}
\begin{IEEEproof}
See Appendix \ref{App:ProofCDFHarvEng}.
\end{IEEEproof}
\begin{remark}\label{Rem:Remark1}
The derivations of the tight lower bounds in \eqref{Eqn:LapTransHavEng},  \eqref{Eqn:CDFHarvEng} and \eqref{Eqn:CDFHarEngAlp4} are attributed to the fact that the locations between all the non-void BSs are very much weakly correlated \cite{CHLLCW1502,CHLLCW16}. These three lowers bounds will asymptotically become exact as the cell load of each tier goes to infinity (i.e., $q_m\rightarrow 1$ as $\ell_m\rightarrow\infty$) since higher cell loads give rise to weaker location correlations between the non-void BSs. 
\end{remark}
\begin{remark}
Although $\Psi(\theta,u)$ in \eqref{Eqn:CDFHarvEng} cannot be found in closed form for $\alpha\neq 4$, we can resort to numerical techniques to evaluate it. In other words, we can still evaluate $F_{P_{eh}}(\theta)$ in \eqref{Eqn:CDFHarvEng} for any values of $\alpha$ numerically.  
\end{remark}
According to Theorem \ref{Thm:CDFHarvEng}, we realize that the cell load of each tier essentially influences the non-void probability of the BSs in each tier so as to remarkably impact the accuracy of the distribution of the harvested power, especially when the HetNet is densely deployed with many BSs. The results in Theorem \ref{Thm:CDFHarvEng} can provides us with some insights into how the statistical properties of the harvested power are generally characterized by user association, transmit beamforming and cell load of each tier.  To further expound and demonstrate how the cell loads impact the CDF of $P_{eh}$, some specific examples and applications of Theorem \ref{Thm:CDFHarvEng} are elaborated as follows. 

\subsubsection{The Generality of the Expression of $F_{P_{eh}}(\theta)$}
The expression of $F_{P_{eh}}(\theta)$ can be used for any particular user association schemes covered by the GUA scheme in \eqref{Eqn:UserAssScheme}.  For instance, the CDF of $P_{eh}$ in \eqref{Eqn:CDFHarEngAlp4} for the MRPA scheme with $w_m=P_m$ in \eqref{Eqn:UserAssScheme} can be easily found as
\begin{align}\label{Eqn:PowerHarvMRPA}
F_{P_{eh}}(\theta) \gtrapprox & \sum_{m=1}^{M} \frac{2\vartheta_m}{\sqrt{\pi}}\bigintssss_{0}^{\infty} \left(1+\frac{\pi^2\lambda^2_{\Sigma}\eta(1-\rho)}{4N_m\theta v^2}\right)^{-N_m}\times\nonumber\\
&e^{-\left(\frac{\pi\lambda_{\Sigma}}{2\sqrt{\theta}}\sum_{m=1}^{M}\phi_m(v)+v\right)^2}\dif v,
\end{align}
where  $\vartheta_m=\sqrt{P_m}\lambda_m/\lambda_{\Sigma}$, $\lambda_{\Sigma}=\sum_{k=1}^{M}\sqrt{P}_k\lambda_k$, and $\phi_m(v)$ is
\begin{align*}
\phi_m(v) = & \vartheta_mq_m\sqrt{\eta(1-\rho)}\times\nonumber\\ &\left[\frac{\pi}{2}- \tan^{-1}\left(\frac{2v}{\pi\lambda_{\Sigma}}\sqrt{\frac{\theta }{\eta(1-\rho)}}\right)\right].
\end{align*}
From the received power point of view, users should adopt the MRPA scheme to associate their BS in order to maximize their received power. Hence, if the MRPA scheme is adopted and the cell load of each tier goes to infinity (i.e. $\ell_m\rightarrow\infty$), no void BSs exist in the HetNet and users thus can receive the maximum power. In this case, \eqref{Eqn:PowerHarvMRPA} reduces to the exact result given by
\begin{align}\label{Eqn:CDFHavEngLargeCellLoad}
\lim_{\stackrel{\ell_m\rightarrow\infty}{m\in\mathcal{M}}}F_{P_{eh}}(\theta) = &\sum_{m=1}^{M} \frac{2\vartheta_m}{\sqrt{\pi}}\bigintssss_{0}^{\infty} \left(1+\frac{\pi^2\lambda^2_{\Sigma}\eta(1-\rho)}{4N_m\theta v^2}\right)^{-N_m}\times \nonumber\\
&e^{-\left(\frac{\pi\lambda_{\Sigma}}{2\sqrt{\theta}}\sum_{m=1}^{M}\phi_m(v)+v\right)^2}\dif v
\end{align}
\textit{which is the CDF of the received maximum power that can be harvested for the case of the tier-$m$ BSs with $N_m$ antennas and path-loss exponent $\alpha=4$}. 

\subsubsection{The Lowest Limit of the CDF of the Harvested Power} As pointed out above,   increasing the cell load of each tier is able to increase the received power because the intensity of non-void BSs increases and more interference powers are generated. Another way to significantly increase the received power is to use a large-scale antenna array at each BS. In other words, the received power of users using the MRPA scheme maximizes when the cell load of each tier and the number of transmit antennas of each BS both go to infinity. In this case, the upper limit of the CDF of the wireless transfered power can be achieved. According to \eqref{Eqn:CDFHavEngLargeCellLoad}, for example, as $N_m$ goes infinity for all $m\in\mathcal{M}$, we further have
\begin{align}\label{Eqn:LimitCDFPowTrans}
\lim_{\stackrel{N_m,\ell_m\rightarrow\infty}{m\in\mathcal{M}}}F_{P_{eh}}(\theta) = & \frac{2}{\sqrt{\pi}}\bigintsss_{0}^{\infty} e^{-\left[\frac{\pi^2\lambda^2_{\Sigma}\eta(1-\rho)}{4\theta v^2}\right]}\times\nonumber\\
& e^{-\left(\frac{\pi\lambda_{\Sigma}}{2\sqrt{\theta}}\sum_{m=1}^{M}\phi_m(v)+v\right)^2}\dif v,
\end{align}
which is \textit{the fundamental lowest limit of the CDF of the harvested power for the case of $\alpha=4$}. We can use \eqref{Eqn:CDFHarvEng} to find the other upper limits of the CDF of the harvested power for any other values of $\alpha$. Moreover, we can show that the limit in \eqref{Eqn:LimitCDFPowTrans} is well approximated by
\begin{align}
\lim_{\stackrel{N_m,\ell_m\rightarrow\infty}{m\in\mathcal{M}}}F_{P_{eh}}(\theta) \sim  \exp\left(-2\pi\lambda_{\Sigma}\sqrt{\frac{\eta(1-\rho)}{\theta}}\right)
\end{align}
if $\frac{\lambda_{\Sigma}}{\sqrt{\theta}}\ll 1$, which indicates $P_{eh}$ has a heavy-tailed distribution, especially as the HetNet is not very dense. 

\subsubsection{The Outage of Energy Harvesting}
If $\theta$ is set as $\underline{P_{eh}}$ that is the minimum power required to activate the energy harvesting circuit, the outage probability of energy harvesting at a user is defined as
\begin{align}\label{Eqn:OutProbEngHarv}
\epsilon_{eh}\defn \mathbb{P}\left[P_{eh}<\underline{P_{eh}}\right]=F_{P_{eh}}(\underline{P_{eh}}).
\end{align}
To make $\epsilon_{eh}$ small, \eqref{Eqn:CDFHarvEng} and \eqref{Eqn:CDFHarEngAlp4} both indicate that $\lambda^2_{\Sigma}(1-\rho)$ has to be sufficiently large, which means deploying more BSs and allocating more power for energy harvesting are the two effective methods of suppressing the outage probability of energy harvesting. Since allocating more power for energy harvesting definitely affects the SINR performance of information decoding and $F_{P_{eh}}(\theta)$ is dominated by $\lambda^2_{\Sigma}$, the most efficient method of reducing $\epsilon_{eh}$ is to deploy more BSs. In addition, for $\lambda_{\Sigma}/\sqrt{\underline{P_{eh}}}\ll 1$, $\epsilon_{eh}$ exponentially decreases as the BS intensities increase since $P_{eh}$ has a heavy-tailed distribution. In addition, $F_{P_{eh}}(\underline{P_{eh}})$ in \eqref{Eqn:OutProbEngHarv} can be applied to find the minimum value of $\lambda_{\Sigma}$ that needs to achieve some specific lower bond on $\epsilon_{eh}$ (i.e., we are able to know how to densely deploy the BSs in the HetNet so as to satisfy the minimum requirement on the outage probability of energy harvesting.)

\subsubsection{The Outage of Self-Powered Sustainability} 
According to \eqref{Eqn:HavestedPower}, the energy harvested by each user can be explicitly expressed as
\begin{align}\label{Eqn:HarvestedEnergy}
E_{eh} = (\beta\tau) P_{eh}= \eta (1-\rho)\beta\tau P_{dl},
\end{align}
whereas each user needs the energy of $\int_{0}^{(1-\beta)\tau}Q(t)\dif t$ for its uplink transmission. Due to random channel variations, the harvested energy $E_{eh}$ may not be always larger than the energy needed for uplink transmission. Hence, here raises the concept of the outage of ``self-powered sustainability'', which refers to the outage scenario in which the energy harvested by users is not enough to completely support the energy needed for the uplink transmission of users. As such, the outage probability of self-powered sustainability of users is defined as
\begin{align}\label{Eqn:OutProbSelfPower}
\epsilon_{ps}& \defn \mathbb{P}\left[E_{eh}<\int_{0}^{(1-\beta)\tau}Q(t)\dif t\right]\nonumber\\
&=F_{P_{eh}}\left(\frac{\int_{0}^{(1-\beta)\tau}Q(t)\dif t}{\beta\tau}\right),
\end{align}   
which can be further simplified as $F_{P_{eh}}((1-\beta)Q/\beta)$ for constant transmit power $Q$. In other words, $\epsilon_{ps}$ can be readily obtained by using \eqref{Eqn:CDFHarvEng} with $\theta=\int_{0}^{(1-\beta)\tau}Q(t)\dif t/\beta\tau$ (or $\theta=(1-\beta)Q/\beta$ if $Q$ is a constant).  A low value of $\epsilon_{ps}$ indicates that users may not need to replenish their battery energy from external power sources very often. In Section \ref{Sec:LinkRateEnergyEff}, we will study the energy efficiency of the proposed power-splitting SWIPT model and how to maximize with the constraint on the outage probability of self-powered sustainability of users. 

\subsection{The Average Harvested Energy and Its Implications} \label{SubSec:AvgHarEng}
The average of the harvested energy $E_{eh}$ in \eqref{Eqn:HarvestedEnergy} can be explicitly found as shown in the following theorem.
\begin{theorem}\label{Thm:MeanHarvEng}
If the GUA scheme in \eqref{Eqn:UserAssScheme} is adopted by all users in the HetNet, then the average of $E_{eh}$ in \eqref{Eqn:HarvestedEnergy} can be explicitly found as
\begin{align}\label{Eqn:MeanHarvEng}
\mathbb{E}[E_{eh}]=&\beta \eta(1-\rho)\tau\sum_{m=1}^{M} \vartheta_m\frac{P_m}{w_m}\bigg\{(\pi\lambda_{\Sigma})^{\frac{\alpha}{2}}\bigg[\Gamma\left(1-\frac{\alpha}{2},\pi \lambda_{\Sigma}\right)\nonumber\\
&-q_m\vartheta^{\frac{\alpha}{2}-1}_m\Gamma\left(1-\frac{\alpha}{2},\pi \lambda_{\Sigma}\vartheta_m\right)\bigg]+\left(\frac{2}{\alpha-2}\right)q_m\times\nonumber\\
&(\pi\lambda_{\Sigma})e^{-\pi\lambda_{\Sigma}\vartheta_m}\bigg\}.
\end{align}
\end{theorem}
\begin{IEEEproof}
See Appendix \ref{App:ProofMeanHarvEng}.
\end{IEEEproof}
The average harvested energy in Theorem \ref{Thm:MeanHarvEng} is a  general expression for any user association scheme specified in \eqref{Eqn:UserAssScheme}.  The MRPA scheme, for example, achieves the average harvested energy given by
\begin{align}\label{Eqn:AvgHarEngMRPA}
\mathbb{E}[E_{eh}]=&\beta \eta(1-\rho)\tau\sum_{m=1}^{M} \vartheta_m\bigg\{(\pi\lambda_{\Sigma})^{\frac{\alpha}{2}}\bigg[\Gamma\left(1-\frac{\alpha}{2},\pi \lambda_{\Sigma}\right)\nonumber\\
&-q_m\vartheta^{\frac{\alpha}{2}-1}_m\Gamma\left(1-\frac{\alpha}{2},\pi \lambda_{\Sigma}\vartheta_m\right)\bigg]+\left(\frac{2}{\alpha-2}\right)q_m\times\nonumber\\
&(\pi\lambda_{\Sigma})e^{-\pi\lambda_{\Sigma}\vartheta_m}\bigg\},
\end{align}
where $\lambda_{\Sigma}=\sum_{k=1}^{M}P^{\frac{2}{\alpha}}_k\lambda_k$ and $\vartheta_m=P^{\frac{2}{\alpha}}_m\lambda_m/\lambda_{\Sigma}$. Mostly importantly, it reveals how the  non-void cell probabilities (or cell loads) impact the average harvested energy.  Other important implications learned from \eqref{Eqn:MeanHarvEng} are elaborated in the following.
 
\subsubsection{The Asymptotic Properties of $\mathbb{E}[E_{eh}]$} In practice, the intensities of the BSs are usually fairly small, i.e., $\lambda_{m}\ll 1$ for all $m\in\mathcal{M}$. Accordingly, as $\lambda_{\Sigma}$ is very small, $\mathbb{E}[E_{eh}]$ in \eqref{Eqn:MeanHarvEng}  can be accurately approximated as
\begin{align}
\mathbb{E}[E_{eh}]\approx \beta \eta(1-\rho)\tau\left(\frac{2\pi}{\alpha-2}\right)\lambda_{\Sigma}\sum_{m=1}^{M} \frac{P_m}{w_m}\vartheta_mq_m
\end{align}
since $e^{-\pi\lambda_{\Sigma}\vartheta_m}\approx 1$. Namely, the average harvested energy is proportional to $\lambda_{\Sigma}\sum_{m=1}^{M} \frac{P_m}{w_m}\vartheta_mq_m$, which manifests that $\vartheta_m$, $P_m$ and $q_m$ (or $\ell_m$) are the three key parameters dominating the average harvested energy in a HetNet with moderate BS intensities. For a ultra dense HetNet with large BS intensities, i.e., $\lambda_{m}\gg 1$, we can have 
\begin{align}
\mathbb{E}[E_{eh}]\approx \beta \eta(1-\rho)\tau\left(\frac{2\pi}{\alpha-2}\right)\lambda_{\Sigma}\sum_{m=1}^{M} \vartheta_m\frac{P_m}{w_m}
\end{align}
since $q_m\ll 1$ and $\pi\lambda_{\Sigma} e^{-\pi\lambda_{\Sigma}\vartheta_m}\ll 1$ for all $m\in\mathcal{M}$. Namely, the majority of the harvested energy is from the BS associated with the user since most of BSs in this case are void. 

\subsubsection{The Cell Load Impact on $\mathbb{E}[E_{eh}]$} As can be inferred from \eqref{Eqn:MeanHarvEng}, the average harvested energy is dependable upon the parameters $\vartheta_m$ and $q_m$ that are the function of the tier-$m$ cell load so that it is significantly affected by all the $M$ cell loads. The average harvested energy essentially increases as the cell loads increase since larger cell loads give rise to more actively working BSs so that more powers can be received and harvested by users.  Hence, the largest average harvested energy, which is attained by adopting the MRPA scheme and making the $M$ cell loads go to infinity, is given by 
\begin{align}
\mathbb{E}[E_{eh}]=&\beta \eta(1-\rho)\tau\sum_{m=1}^{M} \vartheta_m\bigg\{(\pi\lambda_{\Sigma})^{\frac{\alpha}{2}}\bigg[\Gamma\left(1-\frac{\alpha}{2},\pi \lambda_{\Sigma}\right)-\nonumber\\
&\vartheta^{\frac{\alpha}{2}-1}_m\Gamma\left(1-\frac{\alpha}{2},\pi \lambda_{\Sigma}\vartheta_m\right)\bigg]+\left(\frac{2}{\alpha-2}\right)(\pi\lambda_{\Sigma})\times\nonumber\\
&e^{-\pi\lambda_{\Sigma}\vartheta_m}\bigg\}\nonumber\\
\approx &\beta \eta(1-\rho)\tau\left(\frac{2\pi}{\alpha-2}\right)\lambda_{\Sigma}\sum_{m=1}^{M} \frac{P_m}{w_m}\vartheta_m,
\end{align}
where the approximated result is obtained by considering $\lambda_{\Sigma}\ll 1$. Note that $\mathbb{E}[E_{eh}]$ is essentially impacted by how users associate with their BS since the cell loads highly depend on which user association scheme is adopted in the network.

\subsubsection{The Self-Energy Sustainability of Users} In order to make the harvested energy able to completely support the energy needed for uplink transmission, the following inequality must hold
\begin{align}
\mathbb{E}\left[E_{eh}\right]\geq \int_{0}^{(1-\rho)\tau} Q(t)\dif t,
\end{align}
which can reduce to $\mathbb{E}\left[E_{eh}\right]\geq (1-\rho)\tau Q$ if $Q$ is a constant. If this inequality holds for each user, that means the network has ``self-powered sustainability". According to this inequality and $\mathbb{E}\left[E_{eh}\right]$ in \eqref{Eqn:MeanHarvEng}, we are able to know under which user association scheme how much cell load and BS intensity of each tier are needed to achieve the self-energy sustainability.  Once the self-energy sustainability of users is attained, users would never need to replenish their battery by using external power sources.  In the following subsection, some simulation results are provided to validate our analyses and findings for the harvested power and average harvested energy. 

\subsection{Numerical Examples and Verifications}
\begin{table*}[!t]
	\centering
	\caption{Network Parameters for Simulation\cite{CHLKLF16,GAVGCD11,MDRWL17}}\label{Tab:SimPara}
	\begin{tabular}{|c|c|c|}
		\hline Parameter $\setminus$ BS Type (Tier $m$)& Macrocell (1) & Picocell (2)\\ 
		\hline Power $P_m$ (W) & 40 & 10\\ 
		\hline Intensity $\lambda_m$ (BSs/km$^2$) & $1$ (or see figures) & $50\lambda_1$ (or see figures)    \\ 
		\hline Number of Antennas $N_m$ & 8 & 4 \\ 
		\hline Tier-$m$ User Association Weight $w_m$ & $P_1$ & $P_2$ \\
		\hline Power Splitting Factor $\rho$ &\multicolumn{2}{c|}{0.5} \\ 
		\hline Downlink Time Fraction $\beta$ &\multicolumn{2}{c|}{0.75} \\
		\hline Power Conversion  Efficiency $\eta$ &\multicolumn{2}{c|}{0.85}\\
		\hline Minimum Required Power for Energy Harvesting $\underline{P_{eh}}$ (mW) &\multicolumn{2}{c|}{0.2} \\
		\hline Transmit Power of Users $Q$ (mW) &\multicolumn{2}{c|}{1} \\
		\hline Time Duration of Downlink and Uplink $\tau$ (sec) &\multicolumn{2}{c|}{1} \\
		\hline Path-loss Exponent $\alpha$ & \multicolumn{2}{c|}{2.5} \\ 
		\hline 
	\end{tabular} 
\end{table*}

\begin{figure*}[!t]
	\centering
	\includegraphics[width=\textwidth]{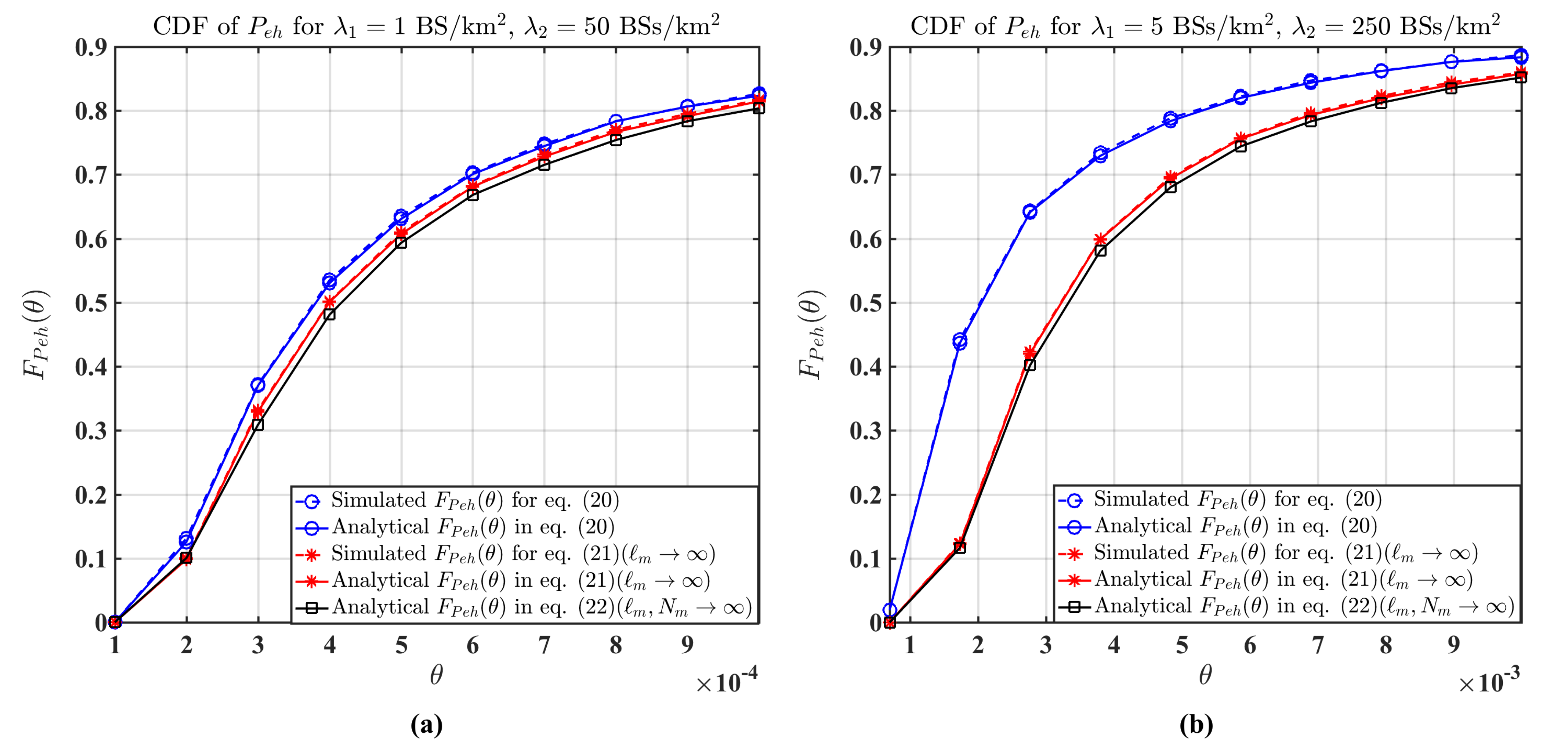}
	\caption{The simulation results of the CDF of $P_{eh}$ in a two-tier HetNet: (a) $\lambda_1=1$ BS/km$^2$ and $\lambda_2=50\lambda_1$, (b) $\lambda_1=5$ BSs/km$^2$ and $\lambda_2=50\lambda_1$}
	\label{Fig:CDFHarvestedPower}
\end{figure*}

In this subsection, we present some numerical results for a two-tier HetNet where the first tier consists of the macro BSs and the second tier consists of the picocell BSs. All users adopt the MRPA scheme to associate with their BS, and all network parameters for simulation are listed in Table \ref{Tab:SimPara}. We first show the simulation results of the CDF of $P_{eh}$ in Fig. \ref{Fig:CDFHarvestedPower} for the case of $\lambda_2=50\lambda_1$. As shown in Fig. \ref{Fig:CDFHarvestedPower}, the simulated results of $F_{P_{eh}}(\theta)$ for the cases in \eqref{Eqn:PowerHarvMRPA} and \eqref{Eqn:CDFHavEngLargeCellLoad} are slightly higher than their analytical results in \eqref{Eqn:PowerHarvMRPA} and \eqref{Eqn:CDFHavEngLargeCellLoad}. Therefore, we have verified that the lower bound found in \eqref{Eqn:CDFHarvEng} is indeed very tight and accurate. Moreover, Fig. \ref{Fig:CDFHarvestedPower} also shows the lowest limit of $F_{P_{eh}}(\theta)$ in \eqref{Eqn:LimitCDFPowTrans} that illustrates an important fact, that is, without modeling the cell load in the received signal power, the CDF of $P_{eh}$ would be close to its fundamental lowest limit, which is essentially not true. There certainly exists a gap between the results in \eqref{Eqn:PowerHarvMRPA} and \eqref{Eqn:LimitCDFPowTrans} and this gap becomes larger and larger as the cell load of each tier is getting smaller and smaller. 

\begin{figure*}[!t]
	\centering
	\includegraphics[width=\textwidth]{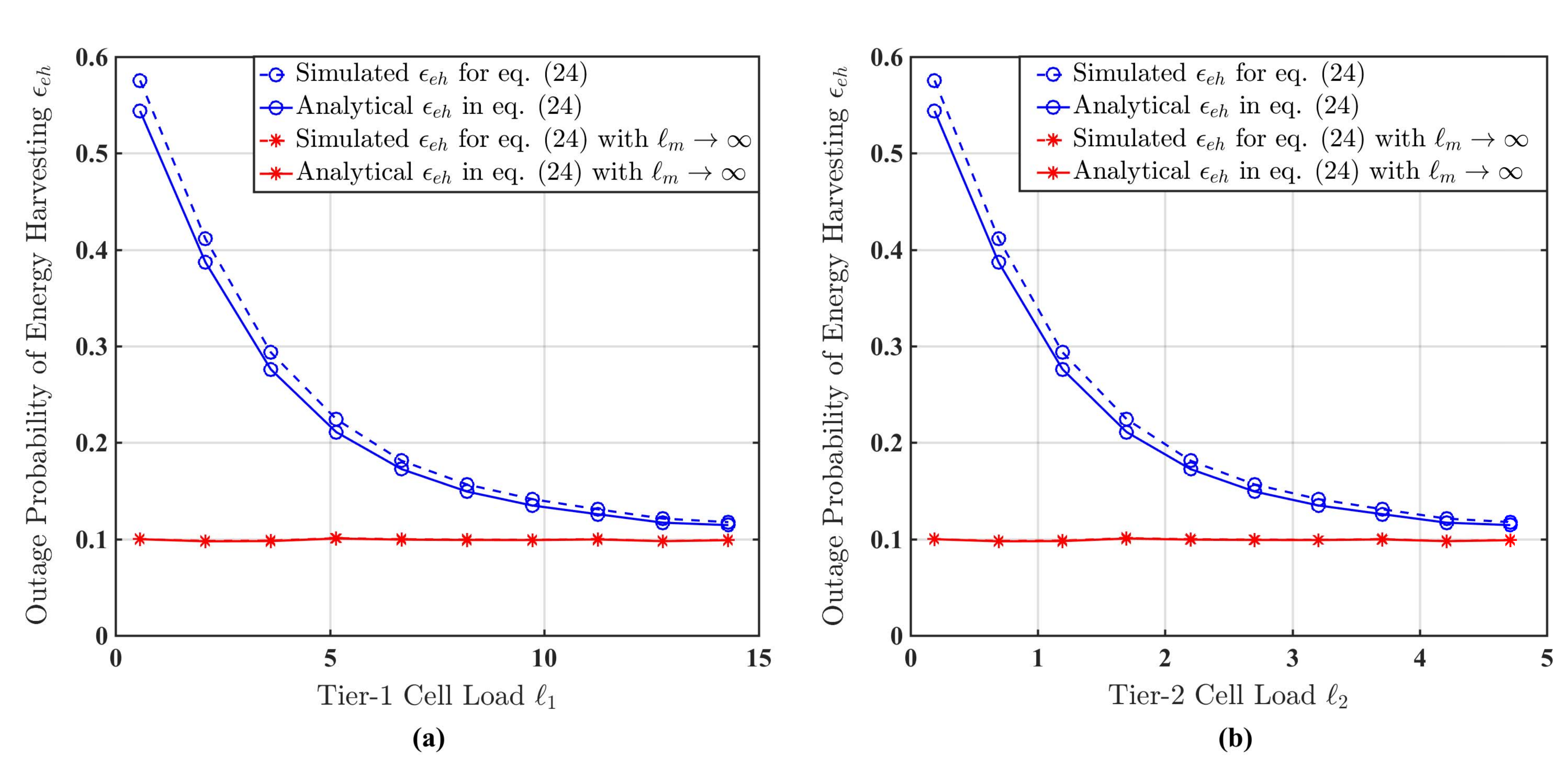}
	\caption{The simulation results of the outage probability $\epsilon_{eh}$ of energy harvesting at a user with $\underline{P_{eh}}=0.2$ mW: (a) the outage probability of energy harvesting for the tier-1 cell load, (b) the outage probability of energy harvesting for the tier-2 cell load.}
	\label{Fig:OutProbEngHarv}
\end{figure*}

\begin{figure*}[!t]
	\centering
	\includegraphics[width=\textwidth]{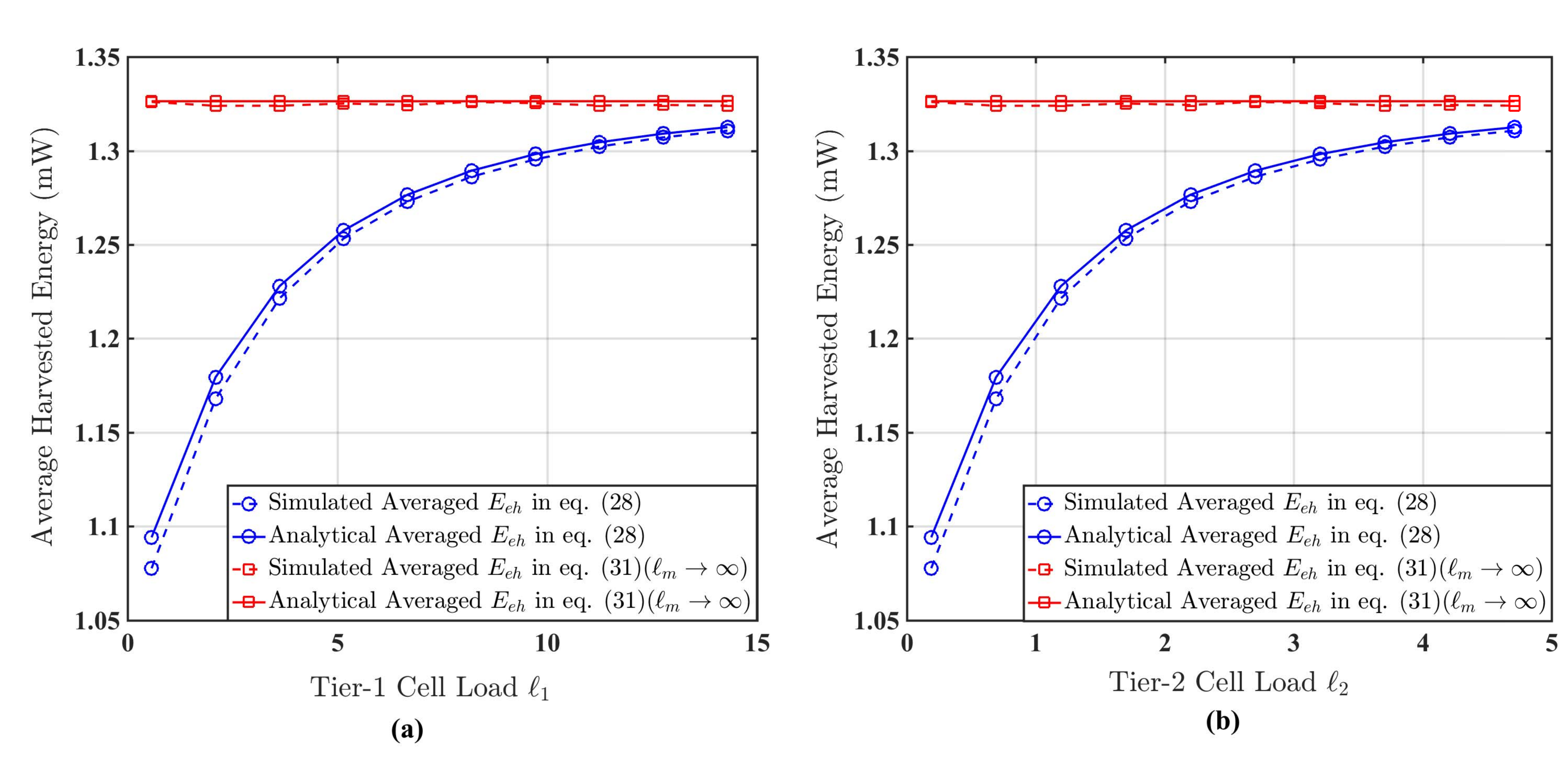}
	\caption{The simulation results of the average harvested energy $\mathbb{E}[E_{eh}]$: (a) $\mathbb{E}[E_{eh}]$ for the tier-1 cell load, (b) $\mathbb{E}[E_{eh}]$ for the tier-2 cell load.}
	\label{Fig:AvgHarEng}
\end{figure*}

The simulation results of the outage probability of energy harvesting at a user are shown in Fig. \ref{Fig:OutProbEngHarv}. As expected, the simulated results of $\epsilon_{ps}$ in the case of finite cell loads (i.e. $\ell_m<\infty$) are slightly larger than their corresponding analytical results in \eqref{Eqn:OutProbSelfPower} so that the result in \eqref{Eqn:OutProbSelfPower} found by using \eqref{Eqn:PowerHarvMRPA} is a very accurate and tight bound. Most importantly, they indicate how the cell load of each tier influences them: as the cell loads increase, $\epsilon_{ps}$ decreases and eventually converges to its fundamental lowest limit that is achieved when all BSs in the HetNet are non-void and actively transmitting. This observation manifests a crucial fact that the key to significantly reducing $\epsilon_{eh}$ is not by unilaterally boosting either the user intensity or the BS intensities, but by boosting the cell load of each tier. Next, Fig. \ref{Fig:AvgHarEng} presents the simulation results of the average harvested energy $\mathbb{E}[E_{eh}]$. In Fig. \ref{Fig:AvgHarEng}, we also can observe that the average harvested energy increases as the cell load of each tier increases and they eventually converge to their upper limit as the cell load of each tier goes to infinity. Overall speaking, the analytical results are fairly close to their corresponding simulated results, which verifies the correctness and accuracy of $\mathbb{E}[E_{eh}]$ in \eqref{Eqn:AvgHarEngMRPA}.

\section{Link Rate Analysis and Energy Efficiency Optimization}\label{Sec:LinkRateEnergyEff}
In previous section, we have analyzed the harvested power and energy and now we are interested in how the harvested power and energy affect the performances of the downlink and uplink rates. Our goal here is to get some insights into not only how to split the total received powers for downlink data decoding and energy harvesting but also how to bisect the total transmission duration $\tau$ so that the energy efficiency of a user using SWIPT can be maximized. Before proceeding the following link rate analyses, we need to first specify the signal-to-interference plus noise ratio (SINR) models for a downlink MISO channel and an uplink SIMO channel. First of all, the downlink SINR of the typical user can be defined as 
\begin{align}
\gamma_{dl} \defn \frac{\rho P_*H_*\|B_*\|^{-\alpha}}{\rho I_{dl}+\sigma^2}=\frac{ P_*H_*\|B_*\|^{-\alpha}}{I_{dl}+\sigma^2/\rho},
\end{align}
where $\sigma^2$ is the noise power induced by RF to baseband conversion at the receiver \cite{YASRCYL18}. For the typical user, the uplink SIR at its associated BS can be expressed as\footnote{For simplicity, we consider that the uplink transmission is interference-limited since the thermal noise at the BS is usually very small compared to the uplink interference.}
\begin{align}
\gamma_{ul}\defn \frac{QG_*\|B_*\|^{-\alpha}}{I_{ul}},
\end{align}
where $G_*$ denotes the uplink channel gain for performing receive beamforming at BS $B_*$, $I_{ul}$ is the interference power received by $B_*$ and it is written as
\begin{align}
I_{ul}\defn \sum_{U_j\in\mathcal{U}_s} QG_j\|B_*-U_j\|^{-\alpha},
\end{align} 
where $\mathcal{U}_s\subseteq\mathcal{U}$ is the set of all users that are scheduled to do uplink transmission and $G_j$ is the fading channel gain from user $j$ to BS $B_*$. Note that the intensity of the point process in set $\mathcal{U}_s$ is the same as the intensity of the non-void BSs, which is $\mu_s=\sum_{m=1}^{M}q_m\lambda_m$, because only the non-void BSs can accept uplink transmissions. As shown in the following subsections, $\gamma_{dl}$ and $\gamma_{ul}$ are used to define the downlink and uplink (ergodic) rates, respectively.  

\subsection{Analysis of the Downlink and Uplink Rates}
By considering a capacity-approaching code is used, the achievable (ergodic) downlink and uplink rates (nats/Hz) can be defined as\footnote{Please note that the following two rates, $c_{dl}$ and $c_{ul}$, are the downlink and uplink rates averaged over space so that they may not accurately indicate the rates of a user located at a specific location in the network.}
\begin{align}\label{Eqn:DefnLinkRate}
c_{dl}\defn \mathbb{E}\left[\log(1+\gamma_{dl})\right]\,\text{ and }\, c_{ul}\defn \mathbb{E}\left[\log(1+\gamma_{ul})\right].
\end{align} 
Their explicit tight lower bounds are obtained as shown in the following theorem.
\begin{theorem}\label{Thm:LinkRates}
Suppose the GUA scheme in \eqref{Eqn:UserAssScheme} is adopted and consider the link rates are defined in \eqref{Eqn:DefnLinkRate}. If the downlink MISO channel gain is $H_*\sim\text{Gamma}(N_m,1/N_m)$ due to transmit beamforming, the tight lower bound on the downlink rate $c_{dl}$ can be shown as
\begin{align}\label{Eqn:DownlinkRate}
c_{dl}\gtrapprox &\sum_{m=1}^{M}\vartheta_m \bigintsss_{0^+}^{\infty}\bigintsss_{0}^{\infty}\left[1-\left(1+\frac{sP_m}{w_mN_m}\right)^{-N_m}\right]\times\nonumber\\
&\frac{\pi\lambda_{\Sigma}}{se^{\pi\lambda_{\Sigma}x[1+\sum_{k=1}^{M}\Phi_k(1,s)]+s\frac{\sigma^2}{\rho}x^{\frac{\alpha}{2}}}} \dif x\dif s,
\end{align}
where $\Phi_k(1,s)$ can be found by \eqref{Eqn:FunPhi}.  For the uplink rate $c_{ul}$, if the uplink SIMO channel gain is $G_*\sim\text{Gamma}(N_m,1)$ due to receive beamforming, its tight lower bound can be found as
\begin{align}\label{Eqn:UplinkRate}
c_{ul}\gtrapprox& \sum_{m=1}^{M}\vartheta_m\bigintssss_{0^+}^{\infty}\frac{1}{s} \left[1-\left(1+\frac{sQ}{w_m}\right)^{-N_m}\right]\times\nonumber\\
&\left[1+\sum_{k=1}^{M}w^{-\frac{2}{\alpha}}_k\Phi_k\left(1,\frac{sQw_k}{\eta(1-\rho)P_k}\right)\right]^{-1}\dif s
\end{align}
in which $\Phi_m(1,\cdot)$ can also be evaluated by using \eqref{Eqn:FunPhi}.
\end{theorem}
\begin{IEEEproof}
See Appendix \ref{App:ProofLinkRates}.
\end{IEEEproof}

\begin{remark}
Note that the tight lower bounds in \eqref{Eqn:DownlinkRate} and \eqref{Eqn:UplinkRate} are the function of all the $M$ cell loads (due to $\vartheta_m$ and $\Phi_m(\cdot,\cdot)$ for all $m\in\mathcal{M}$) and they are derived based on the assumption that all non-void BSs that are actually weakly correlated still form a homogeneous PPP. These two lower bounds will become exact as the cell load of each tier goes to infinity since no void BSs almost surely exist in the network under this situation.   
\end{remark}
\begin{remark}
For $\sigma^2=0$, the result in \eqref{Eqn:DownlinkRate} simply reduces to
\begin{align}
c_{dl}\gtrapprox& \sum_{m=1}^{M}\vartheta_m\bigintsss_{0^+}^{\infty}\left[1-\left(1+\frac{sP_m}{w_mN_m}\right)^{-N_m}\right]\times\nonumber\\
&\frac{\dif s}{s[1+\sum_{k=1}^{M}\Phi_k(1,s)]}.
\end{align}
Whereas for $\alpha=4$, the result in \eqref{Eqn:DownlinkRate} can be further simplified as
\begin{align}\label{Eqn:DownlinkRatewithAlpha4}
\hspace{-0.2in}c_{dl}\gtrapprox & \sum_{m=1}^{M}\vartheta_m\int_{0^+}^{\infty}\frac{\pi^{\frac{3}{2}}\lambda_{\Sigma}\sqrt{\rho}}{2\sigma s^{\frac{3}{2}}}\left[1-\left(1+\frac{sP_m}{w_mN_m}\right)^{-N_m}\right]\times\nonumber\\
&\mathrm{erfcx}\left(\frac{\pi\lambda_{\Sigma}[1+\sum_{k=1}^{M}\Phi_k(1,s)]}{2\sqrt{s\sigma^2/\rho}}\right)\dif s,
\end{align}
where $\mathrm{erfcx}(x)\defn e^{x^2}\mathrm{erfc}(x)$ and $\mathrm{erfc}(x)=1-\frac{2}{\sqrt{\pi}}\int_{0}^{x} e^{-t^2} \dif t$ is the complementary error function. 
\end{remark}

The results in Theorem \ref{Thm:LinkRates} are very general and have a couple of implications that are worth mentioning in the following. First, the fundamental upper limits of the downlink and uplink rates can be achieved by adopting the MRPA scheme and using a large-scale antenna array at the BS. Namely, letting $w_m=P_m$ and $N_m\rightarrow\infty$, we are going to have
\begin{align}
\lim_{N_m\rightarrow\infty}c_{dl}\gtrapprox \bigintssss_{0^+}^{\infty}\bigintssss_{0}^{\infty}\frac{\pi\lambda_{\Sigma}\left(1-e^{-s}\right)}{se^{\pi\lambda_{\Sigma}x[1+\sum_{k=1}^{M}\Phi_k(1,s)]+s\frac{\sigma^2}{\rho}x^{\frac{\alpha}{2}}}} \dif x\dif s
\end{align} 
and 
\begin{align}
\lim_{N_m\rightarrow\infty}c_{ul}\gtrapprox \bigintsss_{0^+}^{\infty}\frac{\dif s}{s\left[1+\sum_{k=1}^{M}P^{-\frac{2}{\alpha}}_k\Phi_k\left(1,\frac{sQ}{\eta(1-\rho)}\right)\right]}.
\end{align}
To the best of our knowledge, these two upper limits on the downlink and uplink rates with cell load modeling  have not been found in the literature. Next, we should notice that $c_{dl}$ is significantly affected by the power splitting factor $\rho$: more received power for energy harvesting gives rise to less downlink rate and vice versa. Thus, there exists a fundamental trade-off between data transmission and energy harvesting, whereas such a trade-off  can be mitigated by maintaining the cell load of each tier below some proper value. In addition, the explicit results of the rates shown in Theorem \ref{Thm:LinkRates} are able to help us define the energy efficiency of a user using SWIPT and then quantitatively evaluate it, as shown in the following subsection. Accurately evaluating the energy efficiency certainly benefits our knowledge regarding how to design a high-performance receiver with energy harvesting.

\subsection{Analysis and Optimization of the Energy Efficiency} 
For a HetNet with SWIPT in the downlink, the energy efficiency of a user is defined as the average sum spectrum efficiency that can be totally transported in downlink and uplink by using one unit of the sum average energy in downlink and uplink. Mathematically, it can be expressed with the unit of (bits/joule) as follows:
\begin{align}
\zeta &\defn \frac{(\beta\tau)\mathbb{E}[\log_2(1+\gamma_{dl})]+(1-\beta)\tau \mathbb{E}[\log_2(1+\gamma_{ul})]}{\mathbb{E}[\beta\tau(P_*+P_{*,on})+(1-\beta)\tau (Q+P_{*,on})]}\\
&=\frac{\beta c_{dl}+(1-\beta) c_{ul}}{\log(2)[\beta\sum_{m=1}^{M}\vartheta_mP_m+ \sum_{m=1}^{M}\vartheta_mP_{m,on}+(1-\beta) Q]}\label{Eqn:DefnEnergyEfficiency}
\end{align}
in which $P_{*,on}\in\{P_{1,on},\ldots,P_{M,on}\}$ stands for the hardware power consumption of BS $B_*$ and $P_{m,on}$ is the  hardware power consumption for an active tier-$m$ BS. The tight lower bound on $\zeta$ can be explicitly found by substituting \eqref{Eqn:DefnLinkRate} and \eqref{Eqn:UplinkRate} into \eqref{Eqn:DefnEnergyEfficiency}. Since $c_{dl}$ is the function of $\rho$ and the average harvested energy is the function of $\beta$, the energy efficiency $\zeta$ apparently depends upon parameters $\rho$ and $\beta$.  As such, our interest now lies in how to optimize parameters $\rho$ and $\beta$ so as to maximize the energy efficiency under some constraints imposed on these two parameters. 

To maintain the self-powered sustainability of users and limit the outage probability of energy harvesting, we formulate the following problem of optimizing the energy efficiency over parameters $\rho$ and $\beta$:
\begin{align}\label{Eqn:OptProbEngEff}
\begin{cases}
\max_{\rho,\beta}& \frac{\beta (c_{dl}-c_{ul})+ c_{ul}}{\beta\sum_{m=1}^{M}\vartheta_m(P_m-Q)+ \sum_{m=1}^{M}\vartheta_m P_{m,on}+Q}\\
\text{s.t.}& \quad (\rho,\beta)\in\mathcal{S}_{\rho,\beta}
\end{cases},
\end{align}
where set $\mathcal{S}_{\rho,\beta}\defn\{(\rho,\beta)\in [\underline{\rho},1)\times(0,\overline{\beta}]:   \mathbb{E}[E_{eh}]\geq (1-\beta)\tau Q, 1>\overline{\epsilon_{eh}}\geq \epsilon_{eh}\}$, $\overline{\epsilon_{eh}}$ is the predesignated upper threshold for the outage probability of energy harvesting and $\overline{\epsilon_{eh}}\geq \epsilon_{eh}$ thus denotes the constraint on  the outage probability of energy harvesting, $\overline{\beta}<1$ is the upper bound on $\beta$, $\underline{\rho}$ is the lower bound on $\rho$, and $\mathbb{E}[E_{eh}]\geq(1-\beta)\tau Q$ is the condition for the self-powered sustainability of users with constant transmit power $Q$ mentioned in Section \ref{SubSec:AvgHarEng}. This optimization problem is feasible if set $\mathcal{S}_{\rho,\beta}\subset [\underline{\rho},1)\times (0,\overline{\beta}]$ is not empty. Once $\mathcal{S}_{\rho,\beta}$  is feasible and the optimal solution  pair ($\rho^{\star}$, $\beta^{\star}$) is found, the receiver of a user is able to not only achieve the maximum energy efficiency, but also maintain the self-powered sustainability and low outage probability of energy harvesting. As a matter of fact, the optimal solution pair ($\rho^{\star}$, $\beta^{\star}$) can be analytically characterized, as summarized in the following theorem.
\begin{theorem}\label{Thm:OptSol}
The optimization problem in \eqref{Eqn:OptProbEngEff} has a feasible set $\mathcal{S}_{\rho,\beta}$ given by
\begin{align}\label{Eqn:FeabSetRhoBeta}
\mathcal{S}_{\rho,\beta}=\mathcal{S}_{\rho}\times \mathcal{S}_{\beta},
\end{align}
where $\mathcal{S}_{\rho}\defn \{\rho\in(0,1): \overline{\epsilon_{eh}}\geq F_{P_{eh}}(\underline{P_{eh}})\}\cap \mathcal{S}'_{\rho}\defn \big[\underline{\rho}, 1-\frac{(1-\beta)Q}{\eta\beta\mathbb{E}[P_{eh}]}\big]$ is nonempty and $\mathcal{S}_{\beta}\defn \left[\frac{Q}{Q+\eta\mathbb{E}[P_{eh}]},\overline{\beta}\right]$. Furthermore, let sets $\underline{\mathcal{S}_{\rho}}$ and $\overline{\mathcal{S}_{\rho}}$ be defined as
\begin{align}\label{Eqn:FebSetLowRho}
\underline{\mathcal{S}_{\rho}}\defn\left\{\rho\in\mathcal{S}_{\rho}:c_{dl}<c_{ul}\left(1+\frac{\sum_{m=1}^{M}\vartheta_m(P_m-Q)}{\sum_{m=1}^{M}\vartheta_mP_{m,on}+Q}\right)\right\}
\end{align}
and
\begin{align}\label{Eqn:FebSetUppRho}
\overline{\mathcal{S}_{\rho}}\defn\left\{\rho\in\mathcal{S}_{\rho}:c_{dl}>c_{ul}\left(1+\frac{\sum_{m=1}^{M}\vartheta_m(P_m-Q)}{\sum_{m=1}^{M}\vartheta_mP_{m,on}+Q}\right)\right\},
\end{align}
respectively. If $\underline{\mathcal{S}_{\rho}}$ is nonempty, then the optimal values of $\rho$ and $\beta$ happen at $\rho^{\star}=\inf \underline{\mathcal{S}_{\rho}}$ and $\beta^{\star}= \frac{Q}{Q+\eta\mathbb{E}[P_{eh}]}$, whereas the optimal values of $\rho$ and $\beta$ happen at $\rho^{\star}=\sup \overline{\mathcal{S}_{\rho}}$ and $\beta^{\star}=\overline{\beta}$ if $\overline{\mathcal{S}_{\rho}}$ is nonempty.
 
\end{theorem}
\begin{IEEEproof}
See Appendix \ref{App:ProofOptSol}.
\end{IEEEproof}

Theorem \ref{Thm:OptSol} essentially reveals the fact that there exists a unique optimal solution pair that maximizes the objective function in \eqref{Eqn:OptProbEngEff} as long as set $\mathcal{S}_{\rho,\beta}$ in \eqref{Eqn:FeabSetRhoBeta} is nonempty and either $\underline{\mathcal{S}_{\rho}}$ in \eqref{Eqn:FebSetLowRho} or $\overline{\mathcal{S}_{\rho}}$ in \eqref{Eqn:FebSetUppRho} is nonempty. This fact is very useful for us to know if it is possible to maximize the energy-efficient performance of a receiver with energy harvesting in the current network deployment and cell load statuses. In the following subsection, we will numerically illustrate the findings in Theorem \ref{Thm:OptSol}. 

\subsection{Simulation Results and Discussions}
In this subsection, the simulation results of the link rates whose tight lower bounds are found in Theorem \ref{Thm:LinkRates} are first presented, which will illustrate whether the bounds found in \eqref{Eqn:DownlinkRate} and \eqref{Eqn:UplinkRate} are tight and accurate or not. Afterwards, we would like to show the simulation results of the energy efficiency and demonstrate whether there exists an optimal pair of parameters $\rho$ and $\beta$ that maximizes the energy efficiency. All network parameters for the simulation here are the same as those shown in Table \ref{Tab:SimPara}. 

\begin{figure*}[!t]
	\centering
	\includegraphics[width=\textwidth]{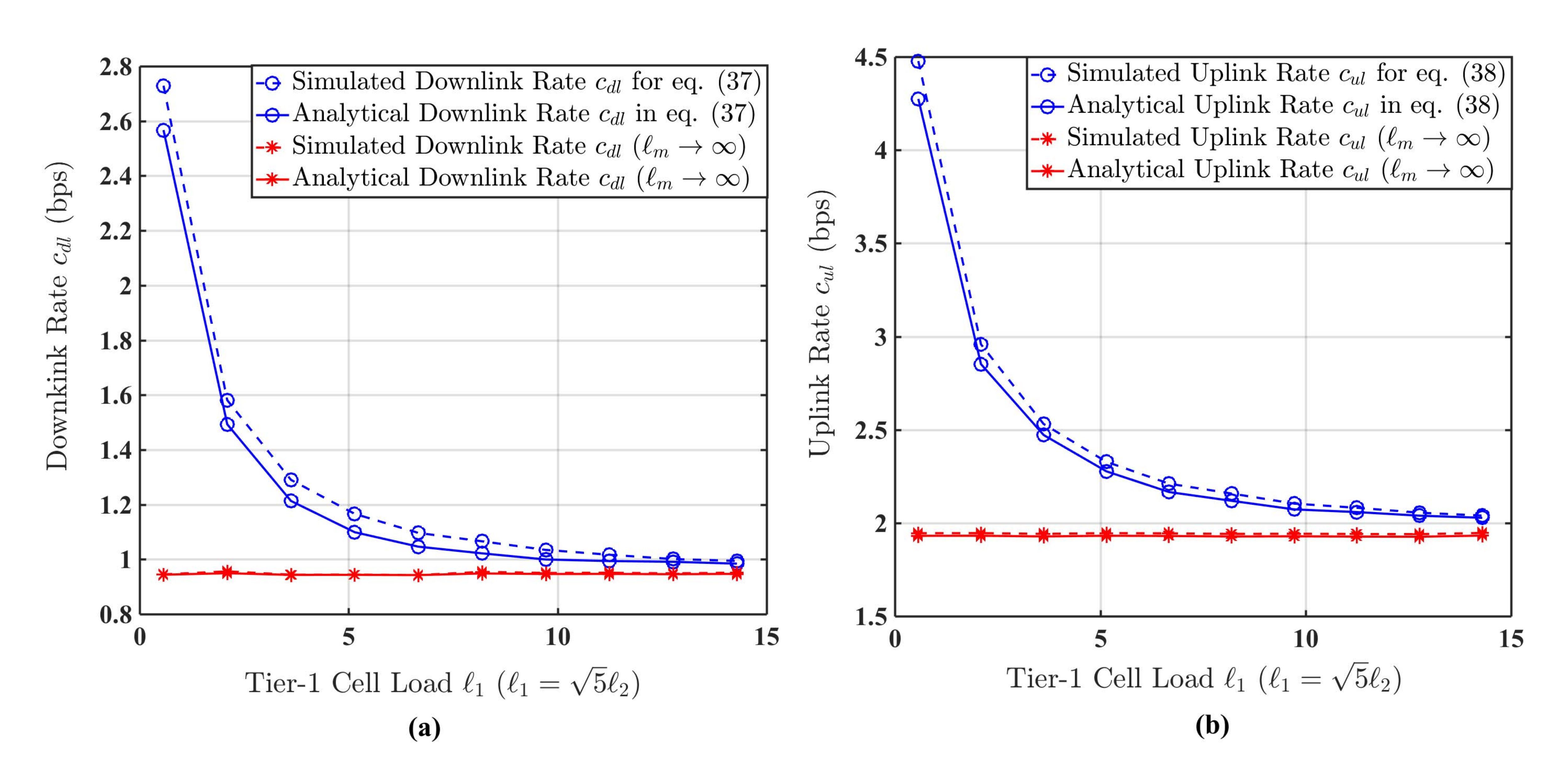}
	\caption{The simulation results of the downlink and uplink rates: (a) downlink rate vs. tier-1 cell load, (b) uplink rate vs. tier-1 cell load.}
	\label{Fig:LinkRates}
\end{figure*}

\begin{figure*}[!t]
	\centering
	\includegraphics[width=\textwidth]{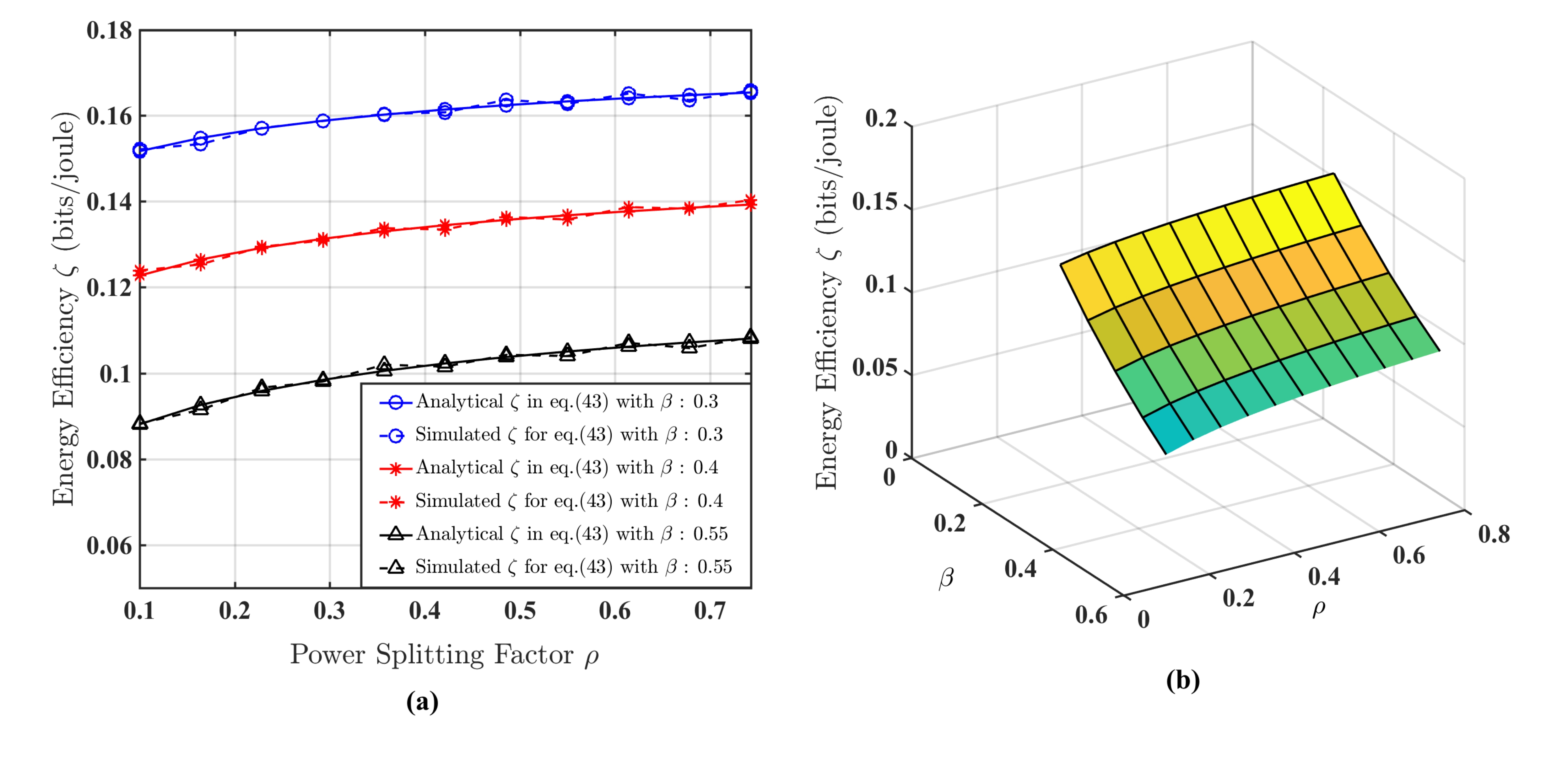}
	\caption{The simulation results of the energy efficiency (bits/joule): (a) energy efficiency $\zeta$ versus power splitting factor $\rho$, (b) the three-dimensional plot of energy efficiency $\zeta$, downlink time fraction $\beta$ and power splitting factor $\rho$. Note that the optimal values of $\rho$ and $\beta$ that maximize the energy efficiency happen at $\rho^{\star}=0.7$ and $\beta^{\star}=0.3$.}
	\label{Fig:EnergyEfficiency}
\end{figure*}

As can be seen in Fig. \ref{Fig:LinkRates}, all the analytical results are just slightly lower than their corresponding simulated results; thereby, the rate expressions in Theorem \ref{Thm:LinkRates} are found correctly and very accurate. Also, we can see that the downlink and uplink rates decrease as the cell load of each tier increases, and this phenomena is owing to the reason that more cell loads bring more non-void BSs and thus induces more interferences in the network. All rates converge to their lowest limits that are obtained by letting all cell loads go to infinity. These rate simulation results verify how significantly the cell loads impact the link rate performances. Without considering the cell load effect in the modeling and analysis, the analyses of the link rates would not be accurate at all. For example, if $\ell_1=2$, the downlink rate without considering cell loads is only about 0.9 bps whereas the downlink rate with considering cell loads is about 1.5 bps, which is about 67\% increase by comparing with 0.9.

The simulation results of the energy efficiency in \eqref{Eqn:DefnEnergyEfficiency} are shown in Fig. \ref{Fig:EnergyEfficiency} for $P_{1,on}=118.7$ W, $P_{2,on}=6.8$ W \cite{GAVGCD11,CHLKLF16}, $\lambda_1=1$ BSs/km$^2$ and $\lambda_2=50$ BSs/km$^2$. We also consider $\beta\in\mathcal{S}_{\beta}$ where $\mathcal{S}_{\beta}$ can be found as $\mathcal{S}_{\beta}=[0.3,1]$ for the current simulation setting. For this simulation setting, set $\underline{\mathcal{S}_{\rho}}$ can be shown as $\underline{\mathcal{S}_{\rho}}=\{\rho\in\mathcal{S}_{\rho}:c_{dl}<3.83\}\approx [0.1, (1-\beta)]$. There are three energy efficiency curves in Fig. \ref{Fig:EnergyEfficiency} for $\beta=0.3$, $\beta=0.4$ and $\beta=0.55$, respectively. Since $\underline{\mathcal{S}_{\rho}}$ is not empty, the optimal values of $\rho$ and $\beta$ are $\rho^{\star}=\inf \underline{\mathcal{S}_{\rho}}=0.7$ and $\beta^{\star}=\frac{Q}{Q+\eta\mathbb{E}[P_{eh}]}=0.3$ according to Theorem \ref{Thm:OptSol}. In indeed, we can see that the highest energy efficiency in Fig. \ref{Fig:EnergyEfficiency} is about 0.165 and it occurs when $\rho=0.7$ and $\beta=0.3$. This validates the statements in Theorem \ref{Thm:OptSol} and clarifies why other values of $\rho$ and $\beta$ cannot achieve an energy efficiency higher than 0.165.

\section{Conclusion}\label{Sec:Conclusion}
In this paper, some fundamental performance metrics of simultaneous wireless information and power transmission in a HetNet are thoughtfully studied from a cell load perspective. The motivation of exploiting how the cell loads impact the SWIPT performances in a HetNet is inspired by the user-centric association behavior that leads to the presence of the void BSs in the HetNet and how likely a BS in a specific tier becomes void is dominated by the cell load of the tier. By considering the void BS impacts in the received signal power model, we first succeed to exploit the fundamental relationships between the statistical properties of the harvested power and the cell loads by deriving the Laplace transforms and mean of the harvested power and energy. We then proceed to study how the link rates are affected by the cell loads. All these derived analytical results are shown to be significantly impacted by the cell load of each tier. The fundamental limits of the CDF of the harvested power and link rates are also characterized. Finally, the problem of optimizing the energy efficiency at users with a power-splitting structure is investigated and the optimal solution to this problem is shown to exist and analytically solvable if the derived constraints hold. 

\appendix [Proofs of Theorems]
\subsection{Proof of Theorem \ref{Thm:LapMeanNthShotNoise}}\label{App:ProofLapMeanNthShotNoise}
Since $X_n$ is the $n$th nearest point in $\mathcal{X}$ to the origin, we have $|X_{n+1}|^2=|X_{1}|^2+|X_{n}|^2$ because $|X_{n}|^2$ is the sum of $n$ i.i.d. RVs which have the same distribution as $|X_1|^2\sim \exp(\pi\lambda_{\mathcal{X}})$ \cite{CHLLCW16,MH12}. As a result, we can also have $\|X_{n+1}\|^2=\|X_1\|^2+\|X_n\|^2$ and this leads to the following results:
\begin{align*}
I_{(n)} &= \widehat{W}_n\left(\|X_n\|^2\right)^{-\frac{\alpha}{2}}+\sum_{n:X_{n+1}\in\mathcal{X}}\frac{W_{n+1}}{\left(\|X_1\|^2+\|X_n\|^2\right)^{\frac{\alpha}{2}}}\\
&\stackrel{d}{=}\|X_n\|^{-\alpha}\left[\widehat{W}_n+\sum_{k:X^{\dagger}_k\in\mathcal{X}^{\dagger}}W_k\left(1+\frac{\|X^{\dagger}_k\|^2}{\|X_n\|^2}\right)^{-\frac{\alpha}{2}}\right],
\end{align*}
where $\stackrel{d}{=}$ denotes the equivalence in distribution, $\mathcal{X}^{\dagger}\defn\{X_k^{\dagger}\in\mathbb{R}^2:k\in\mathbb{N}_+\}$ is a homogeneous PPP of intensity $\lambda_{\mathcal{X}}$, and $X^{\dagger}_k$ is the $k$th nearest point in $\mathcal{X}^{\dagger}$ to the origin. Then the Laplace transform of $I_{(n)}$ can be written as
\begin{align*}
&\mathcal{L}_{I_{(n)}}(s) =\\
&\mathbb{E}\left[\exp\left(-\frac{s}{\|X_n\|^{\alpha}}\left[\widehat{W}_n+\sum_{k:X^{\dagger}_k\in\mathcal{X}^{\dagger}}\frac{W_k}{\left(1+\frac{\|X^{\dagger}_k\|^2}{\|X_n\|^2}\right)^{\frac{\alpha}{2}}}\right]\right)\right].
\end{align*}
For given $\|X_n\|^2=x$, we have the following:
\begin{align*}
&\mathcal{L}_{I_{(n)}}(s)\bigg|_{\|X_n\|^2=x}=\\
&\mathbb{E}\left[\exp\left(-\frac{s}{x^{\frac{\alpha}{2}}} \left[\widehat{W}_n+\sum_{k:X^{\dagger}_k\in\mathcal{X}^{\dagger}}\frac{W_k}{\left(1+\frac{\|X^{\dagger}_k\|^2}{x}\right)^{\frac{\alpha}{2}}}\right]\right)\right]\\
\stackrel{(a)}{=}&\mathcal{L}_{\widehat{W}_n}\left(\frac{x}{s^{\frac{\alpha}{2}}}\right)\exp\left(-\pi\lambda_{\mathcal{X}}\int_{0}^{\infty} \mathbb{E}_W\left[1-e^{- \frac{sW}{x^{\frac{\alpha}{2}}\left(1+\frac{r}{x}\right)^{\frac{\alpha}{2}}}}\right]\dif r\right)\\
\stackrel{(b)}{=}&\mathcal{L}_{\widehat{W}_n}\left(\frac{x}{s^{\frac{\alpha}{2}}}\right)\exp\left(-\pi\lambda_{\mathcal{X}}x\int_{1}^{\infty} \mathbb{P}\left[u\leq x^{-1}\left(s \frac{W}{Y}\right)^{\frac{2}{\alpha}}\right]\dif u\right)\\
=&\mathcal{L}_{\widehat{W}_n}\left(\frac{x}{s^{\frac{\alpha}{2}}}\right)e^{-\pi\lambda_{\mathcal{X}}s^{\frac{2}{\alpha}}\Gamma\left(1-\frac{2}{\alpha}\right)\mathbb{E}\left[W^{\frac{2}{\alpha}}\right]}\times\\
&e^{-\pi\lambda_{\mathcal{X}}\left[\int_{0}^{x} \mathcal{L}_W\left(sv^{-\frac{\alpha}{2}}\right)\dif v-x\right]},
\end{align*}
where $(a)$ follows from the probability generating functional (PGFL) of a homogeneous PPP applied to the point process $\mathcal{X}^{\dagger}$ and $(b)$ follows from the assumption of $Y\sim\exp(1)$. Since $|X_n|\sim \text{Gamma}\left(n,1/\pi\lambda_{\mathcal{X}}\right)$, we can have the result in \eqref{Eqn:LapGenNthShotNoise} by using the definition of the path-loss model $\|\cdot\|^{-\alpha}$.

The mean of $I_{(n)}$ can be explicitly expressed as
\begin{align*}
&\mathbb{E}\left[I_{(n)}\right]\\ =&\mathbb{E}\left\{\|X_n\|^{-\alpha}\left[\widehat{W}_n+\sum_{k:X^{\dagger}_k\in\mathcal{X}^{\dagger}}W_k\left(1+\frac{\|X^{\dagger}_k\|^2}{\|X_n\|^2}\right)^{-\frac{\alpha}{2}}\right]\right\}\\
=&\int_{0}^{\infty} x^{-\frac{\alpha}{2}}\left(\mathbb{E}\left[\widehat{W}_n\right]+\mathbb{E}\left[\sum_{k:X^{\dagger}_k\in\mathcal{X}^{\dagger}}\frac{W_k}{\left(1+\frac{\|X^{\dagger}_k\|^2}{x}\right)^{\frac{\alpha}{2}}}\right]\right)\times\\ &f_{\|X_n\|^2}(x)\dif x\\
\stackrel{(c)}{=}&\frac{(\pi\lambda_{\mathcal{X}})^n}{(n-1)!}\int_{1}^{\infty} \left\{\mathbb{E}\left[\widehat{W}_n\right]+\pi\lambda_{\mathcal{X}}\mathbb{E}\left[W\right]\int_{0}^{\infty}\left(1+\frac{r}{x}\right)^{-\frac{\alpha}{2}}\dif r\right\}\\
& \times x^{n-\frac{\alpha}{2}-1} e^{-\pi\lambda_{\mathcal{X}} x}\dif x\\
=&\frac{(\pi\lambda_{\mathcal{X}})^{\frac{\alpha}{2}}}{(n-1)!} \bigg\{ \mathbb{E}\left[\widehat{W}_n\right]\Gamma\left(n-\frac{\alpha}{2},\pi\lambda_{\mathcal{X}}\right)+\frac{2\mathbb{E}\left[W\right]}{(\alpha-2)}\times\\
&\Gamma\left(n-\frac{\alpha}{2}+1,\pi\lambda_{\mathcal{X}}\right)\bigg\},
\end{align*}
where $(c)$ is obtained by using the Campbell theorem for the PPP of $\mathcal{X}^{\dagger}$ and the fact of $|X_n|^2\sim \text{Gamma}(n,1/\pi\lambda_{\mathcal{X}})$. Then applying the property of $\Gamma(a+1,b)=a\Gamma(a,b)+b^a e^{-b}$ in the result of the last equality yields the result in \eqref{Eqn:MeanGenNthShotNoise}.

\subsection{Proof of Theorem \ref{Thm:CDFHarvEng}}\label{App:ProofCDFHarvEng}
According to the proof of Theorem \ref{Thm:LapMeanNthShotNoise}, we know that the expression of $I_{dl}$ in \eqref{Eqn:DownlinkInterference} can be rewritten as
\begin{align*}
I_{dl} &= \sum_{B_{m,i}\in\mathcal{B}\setminus B_*} \frac{P_mV_{m,i}H_{m,i}}{w_m\left(w^{-\frac{2}{\alpha}}_m\|B_{m,i}\|^2\right)^{\frac{\alpha}{2}}}\\
&\stackrel{d}{=}\sum_{B_{m,j}\in\mathcal{B}} \frac{P_mV_{m,j}H_{m,j}}{w_m\left(w^{-\frac{2}{\alpha}}_*\|B_*\|^2+w^{-\frac{2}{\alpha}}_m\|B_{m,j}\|^2\right)^{\frac{\alpha}{2}}}\\
&\stackrel{d}{=}\sum_{\widetilde{B}_{m,j}\in\widetilde{\mathcal{B}}}\frac{P_m}{w_m}V_{m,i}H_{m,i}\left[\|\widetilde{B}_*\|^2+\|\widetilde{B}_{m,j}\|^2\right]^{-\frac{\alpha}{2}},
\end{align*}
where $\stackrel{d}{=}$ denotes the equivalence in distribution, $\|\widetilde{B}_*\|^2=w^{-\frac{2}{\alpha}}_*\|B_*\|^2$, $\|\widetilde{B}_{m,j}\|^2=w^{-\frac{2}{\alpha}}_m\|B_{m,j}\|^2$, $B_{m,j}$ is the $j$th nearest point in $\mathcal{B}$ to the origin, $\widetilde{\mathcal{B}}\defn \bigcup_{m=1}^M \widetilde{\mathcal{B}}_m$, $\widetilde{\mathcal{B}}_m\defn \{\widetilde{B}_{m,j}\in\mathbb{R}^2: m\in\mathcal{M}, j\in\mathbb{N}_+\}$ denotes a homogeneous PPP of intensity $w^{\frac{2}{\alpha}}_m\lambda_m$, and $\widetilde{B}_{m,j}$ is the $j$th nearest point in $\mathcal{\widetilde{B}}_m$ to the origin. Note that $\widetilde{B}_*$ is the nearest point in set $\widetilde{\mathcal{B}}$ to the origin and $\|\widetilde{B}_*\|^2\sim\exp(\pi\lambda_{\Sigma})$. The above equivalent results in distributions are essentially found based on the fact that $w^{-\frac{2}{\alpha}}_m\|B_{m,i}\|^2$ can be expressed as the sum of a certain number of i.i.d. RVs having the same distribution as $\|\widetilde{B}_*\|^2$, as pointed out in \cite{CHLLCW16,CHLKLF16}.


According to \eqref{Eqn:ReceivedDownLinkPower}, \eqref{Eqn:HavestedPower} and the above equivalent expression of $I_{dl}$, the Laplace transform of $P_{eh}$, $\mathcal{L}_{P_{eh}}(s)$, can be explicitly written as
\begin{align*}
&\mathbb{E}\left[e^{-s(\frac{P_*H_*}{\|B_*\|^{\alpha}}+I_{dl})}\right]=\sum_{m=1}^{M}\vartheta_m \mathbb{E}\bigg\{\exp\bigg[-s\bigg(\frac{P_mH_*}{w_m\|\widetilde{B}_*\|^{\alpha}}\\
&+\sum_{\widetilde{B}_{m,j}\in\widetilde{\mathcal{B}}}\frac{P_mV_{m,i}H_{m,i}}{w_m}\left(\|\widetilde{B}_*\|^2+\|\widetilde{B}_{m,j}\|^2\right)^{-\frac{\alpha}{2}}\bigg)\bigg]\bigg\},
\end{align*}
where $\vartheta_m\defn\mathbb{P}[B_*\in\mathcal{B}_m]$ is the probability that a user associates with a tier-$m$ BS. According to the proof of Theorem \ref{Thm:LapMeanNthShotNoise} and our previous work in \cite{CHLKLF16}, we know $\vartheta_m=w^{\frac{2}{\alpha}}_m\lambda_m/\sum_{k=1}^{M}w^{\frac{2}{\alpha}}_k\lambda_k=\ell_m\lambda_m/\mu$. For given $\|\widetilde{B}_*\|^2= x$, we have the following result
\begin{align*}
I_{dl}\stackrel{d}{=}\sum_{\widetilde{B}_{m,j}\in\widetilde{\mathcal{B}}}\frac{P_m}{w_m}V_{m,i}H_{m,i}\left(x+\|\widetilde{B}_{m,j}\|^2\right)^{-\frac{\alpha}{2}}, 
\end{align*}
and we thus ca n have
\begin{align*}
&\mathcal{L}_{I_{dl}}(s)\bigg|_{\|\widetilde{B}_*\|^2=x }=\\
&\mathbb{E}\left[\exp\left(-s\sum_{\widetilde{B}_{m,j}\in\widetilde{\mathcal{B}}}\frac{P_m}{w_m}V_{m,i}H_{m,i}\left(x+\|\widetilde{B}_{m,j}\|^2\right)^{-\frac{\alpha}{2}} \right)\right]\\
\stackrel{(a)}{\gtrapprox}&\prod_{m=1}^{M}\exp\left(-\frac{\pi\lambda_mq_m(sP_m/w_m)^{\frac{2}{\alpha}}}{\text{sinc}(2/\alpha)}\right)\times\\
&\exp\left(\pi\lambda_mq_m\int_{0}^{x}\frac{s(P_m/w_m)}{s(P_m/w_m)+v^{\frac{\alpha}{2}}}\dif v\right)
\end{align*}
\begin{align*}
=&\prod_{m=1}^{M}\exp\bigg[-\pi\lambda_{\Sigma}\vartheta_mq_m\left(\frac{sP_m}{w_m}\right)^{\frac{2}{\alpha}}\bigg(\frac{1}{\text{sinc}(2/\alpha)}-\\
&\int_{0}^{x(\frac{w_m}{sP_m})^{\frac{2}{\alpha}}}\frac{\dif t}{1+t^{\frac{\alpha}{2}}}\bigg)\bigg] 
\stackrel{(b)}{=} e^{-\pi\lambda_{\Sigma}\sum_{m=1}^{M}\Phi_m(x,s)},
\end{align*}
where $(a)$ is first obtained by using the result in \eqref{Eqn:LapGenNthShotNoise} and then the tight lower bound is obtained by considering all $V_{m,i}$'s that actually weakly correlated are independent \cite{DSWKJM13}, and $(b)$ follows from the definition of function $\Phi_m(\cdot,\cdot)$. Using this result readily leads to the following result:
\begin{align*}
\mathcal{L}_{P_{dl}}(s)\bigg|_{\|\widetilde{B}_*\|^2=x}=&\sum_{m=1}^{M}\vartheta_m\mathcal{L}_{H_*}\left(\frac{sP_m}{w_m}x^{-\frac{\alpha}{2}}\right)\times\\
 &\exp\left[-\pi\lambda_{\Sigma}\sum_{m=1}^{M}\Phi_m(x,s)\right] 
\\
\stackrel{(c)}{=}&\sum_{m=1}^{M}\vartheta_m\left(1+\frac{sP_m}{w_mN_mx^{\frac{\alpha}{2}}}\right)^{-N_m}\times\\ &\exp\left[-\pi\lambda_{\Sigma}\sum_{m=1}^{M}\Phi_m(x,s)\right],
\end{align*}
where $(c)$ follows from the Laplace transform of $H_*\sim\text{Gamma}(N_m,1/N_m)$ whenever $B_*\in\Phi_m$. Therefore, it follows that
\begin{align*}
\mathcal{L}_{P_{dl}}(s) =&\int_{0}^{\infty}\pi\lambda_{\Sigma} \sum_{m=1}^{M}\vartheta_m\left(1+\frac{sP_m}{w_mN_mx^{\frac{\alpha}{2}}}\right)^{-N_m}\times\\ &\exp\left[-\pi\lambda_{\Sigma}\sum_{m=1}^{M}\Phi_m(x,s)-\pi\lambda_{\Sigma}x\right]\dif x,
\end{align*}
since $|\widetilde{B}_*|^2\sim\exp(\pi\lambda_{\Sigma})$. The tight lower bounds in \eqref{Eqn:LapTransHavEng} can be acquired by replacing $P_m$ in the above tight lower bound on $\mathcal{L}_{P_{dl}}(s)$ with $\eta(1-\rho)P_m$.

The explicit result on the CDF of $P_{eh}$ can be found as follows
\begin{align*}
F_{P_{eh}}(\theta) =& \mathcal{L}^{-1}\left\{\frac{\mathcal{L}_{P_{eh}}(s)}{s}\right\}(\theta)\\
\approx &\sum_{m=1}^{M}\vartheta_m\int_{0}^{\infty}\pi\lambda_{\Sigma}\left(1+\frac{P_m}{w_mN_mu^{\frac{\alpha}{2}}}\right)^{-N_m}\times\\
&\mathcal{L}^{-1}\left\{s^{\frac{2}{\alpha}-1}e^{-\pi \lambda_{\Sigma} s^{\frac{2}{\alpha}}\left( \sum_{m=1}^{M} \Phi_m(u,1)+u\right)}\right\}(\theta)\,\dif u.
\end{align*}
For $\alpha=4$, we further can have
\begin{align*}
F_{P_{eh}}(\theta) \approx& \sum_{m=1}^{M} \vartheta_m \int_{0}^{\infty}\lambda_{\Sigma}\sqrt{\frac{\pi}{\theta}}\left(1+\frac{\eta(1-\rho)P_m}{w_mN_mu^2}\right)^{-N_m}\times\\
&e^{-\frac{\pi^2\lambda_{\Sigma}^2}{4\theta}\left( \sum_{m=1}^{M} \Phi_m(u,1)+u\right)^2}\dif u\\
\stackrel{(d)}{=}&\sum_{m=1}^{M} \frac{2\vartheta_m}{\sqrt{\pi}}\int_{0}^{\infty} \left(1+\frac{\pi^2\lambda^2_{\Sigma}\eta(1-\rho)P_m}{4w_mN_m\theta v^2}\right)^{-N_m}\times\\
&\exp\left[-\left(\frac{\pi\lambda_{\Sigma}}{2\sqrt{\theta}}\sum_{m=1}^{M}\phi_m(v)+v\right)^2\right]\dif v,
\end{align*}
where $(d)$ is due to the the variable change of $v=\frac{\pi\lambda_{\Sigma}}{2\sqrt{\theta}}u$. Hence, we
obtain the result in \eqref{Eqn:CDFHarEngAlp4}.

\subsection{Proof of Theorem \ref{Thm:MeanHarvEng}}\label{App:ProofMeanHarvEng}

According to \eqref{Eqn:HarvestedEnergy} and the equivalent expression of $I_{dl}$ shown in the proof of Theorem \ref{Thm:CDFHarvEng}, the mean of the harvested energy at each user can be expressed as
\begin{align*}
\mathbb{E}\left[E_{eh}\right] =& \beta \eta (1-\rho)\tau \mathbb{E}\left[P_{dl}\right]=\beta \eta(1-\rho)\tau\mathbb{E}\left[\frac{P_*H_*}{\|B_*\|^{\alpha}}+I_{dl}\right]\\
=&\beta \eta (1-\rho)\tau\mathbb{E}\bigg[\sum_{m=1}^{M}\vartheta_m\left(\frac{P_m}{w_m}\mathbb{E}[H_*]\right)\|\widetilde{B}_*\|^{-\alpha}\\
&+\sum_{\widetilde{B}_{m,j}\in\widetilde{\mathcal{B}}}\frac{P_mV_{m,i}H_{m,i}}{w_m\left[\|\widetilde{B}_*\|^2+\|\widetilde{B}_{m,j}\|^2\right]^{\frac{\alpha}{2}}}\bigg].
\end{align*}
Using the result in \eqref{Eqn:MeanGenNthShotNoise} for $n=1$, $\mathbb{E}[\widehat{W}_1]=\sum_{m=1}^{M}\vartheta_mP_m\mathbb{E}[H_m]/w_m$, $\mathbb{E}[H_*]=1$, $\mathbb{E}[W_m]=\frac{P_m}{w_m}\mathbb{E}[V_m]\mathbb{E}[H_m]$, $\mathbb{E}[H_m]=1$, and $\mathbb{E}[V_m]=q_m$ leads to the following expression
\begin{align*}
\mathbb{E}\left[E_{eh}\right] =& \beta \eta(1-\rho)\tau(\pi\lambda_{\Sigma})^{\frac{\alpha}{2}} \bigg[\sum_{m=1}^{M}\vartheta_m\frac{P_m}{w_m}\Gamma\left(1-\frac{\alpha}{2},\pi \lambda_{\Sigma}\right)\\
&+\sum_{m=1}^{M}\frac{2q_m\vartheta_m^{\frac{\alpha}{2}}P_m}{(\alpha-2)w_m}\Gamma\left(2-\frac{\alpha}{2},\pi \lambda_{\Sigma}\vartheta_m\right)\bigg]\\
=&\beta \eta(1-\rho)\tau(\pi\lambda_{\Sigma})^{\frac{\alpha}{2}}\bigg[\sum_{m=1}^{M} \vartheta_m\frac{P_m}{w_m}\bigg(\Gamma\left(1-\frac{\alpha}{2},\pi \lambda_{\Sigma}\right)\\
&+\frac{2q_m\vartheta^{\frac{\alpha}{2}-1}_m}{\alpha-2}\Gamma\left(2-\frac{\alpha}{2},\pi \lambda_{\Sigma}\vartheta_m\right)\bigg)\bigg],
\end{align*}
which yields the result in \eqref{Eqn:MeanHarvEng}.

\subsection{Proof of Theorem \ref{Thm:LinkRates}}\label{App:ProofLinkRates}
By using the similar integral transformation technique devised in \cite{CHLHCT17}, we know that the downlink rate $c_{dl}$ can be rewritten as
\begin{align*}
c_{dl} =& \int_{0}^{\infty}\int_{0}^{1} e^{-ty} \mathbb{E}\left[\exp\left(-\frac{t}{\gamma_{dl}}\right)\right]\dif y\dif t\\
=&\int_{0}^{\infty}\int_{0}^{1} e^{-ty} \mathbb{E}\left[\exp\left(-\frac{t(I_{dl}+\sigma^2)}{P_*H_*\|B_*\|^{-\alpha}}\right)\right]\dif y\dif t\\
=&\int_{0}^{\infty} \mathbb{E}\left[\left(\frac{1-e^{-t}}{t}\right)\exp\left(-\frac{t(I_{dl}+\sigma^2)}{P_*H_*\|B_*\|^{-\alpha}}\right)\right] \dif t\\
=&\sum_{m=1}^{M}\vartheta_m\int_{0}^{\infty}\left[1-\mathcal{L}_{H_*}\left(\frac{sP_m}{w_m}\right)\right]\mathbb{E}\left[e^{-\frac{s(I_{dl}+\sigma^2)}{\|\widetilde{B}_*\|^{-\alpha}}}\right]  \frac{\dif s}{s}\\
\stackrel{(a)}{=}&\sum_{m=1}^{M}\vartheta_m\int_{0}^{\infty}\left[1-\left(1+\frac{sP_m}{w_mN_m}\right)^{-N_m}\right]\times\\
&\mathbb{E}\left[\exp\left({-s(I_{dl}+\sigma^2)\|\widetilde{B}_*\|^{\alpha}}\right)\right]  \frac{\dif s}{s},
\end{align*}
where $(a)$ follows from the result in the proof of Theorem \ref{Thm:CDFHarvEng} and $I_{dl}$ is equivalently written as
$$I_{dl}\stackrel{d}{=}\sum_{\widetilde{B}_{m,j}\in\widetilde{\mathcal{B}}}\frac{P_m}{w_m}V_{m,i}H_{m,i}\left[\|\widetilde{B}_*\|^2+\|\widetilde{B}_{m,j}\|^2\right]^{-\frac{\alpha}{2}}.$$

For given $\|\widetilde{B}_*\|^2=x$, we further have
\begin{align*}
c_{dl}\big|_{\|\widetilde{B}_*\|^2=x} =&\sum_{m=1}^{M}\vartheta_m \int_{0}^{\infty}\left[1-\left(1+\frac{sP_m}{w_mN_m}\right)^{-N_m}\right]\times\\
& \frac{\mathcal{L}_{I_{dl}}\left(sx^{\frac{\alpha}{2}}\right)}{se^{s(\sigma^2/\rho)x^{\frac{\alpha}{2}}}}\dif s.
\end{align*}
According to the proof of Theorem \ref{Thm:CDFHarvEng}, we know
\begin{align*}
\mathcal{L}_{I_{dl}}\left(sx^{\frac{\alpha}{2}}\right) \big|_{\|\widetilde{B}_*\|^2=x}&\gtrapprox \exp\left[-\pi\lambda_{\Sigma}\sum_{k=1}^{M}\Phi_k\left(x,sx^{\frac{\alpha}{2}}\right)\right]\\
&=\exp\left[-\pi\lambda_{\Sigma}x\sum_{k=1}^{M}\Phi_k\left(1,s\right)\right],
\end{align*}
which indicates the following results:
\begin{align*}
c_{dl} \gtrapprox& \sum_{m=1}^{M}\vartheta_m \int_{0}^{\infty}\left[1-\left(1+\frac{sP_m}{w_mN_m}\right)^{-N_m}\right]\times\\
& \mathbb{E}_{|\widetilde{B}_*|^2}\left[e^{-sw_m(\sigma^2/\rho)|\widetilde{B}_*|^{\alpha}}\mathcal{L}_{I_{dl}}\left(sw_m|\widetilde{B}_*|^{\alpha}\right)\right]\frac{\dif s}{s}\\
=& \sum_{m=1}^{M}\vartheta_m \int_{0}^{\infty}\left[1-\left(1+\frac{sP_m}{w_mN_m}\right)^{-N_m}\right]\times\\
& \mathbb{E}_{|\widetilde{B}_*|^2}\left[e^{-\pi\lambda_{\Sigma}|\widetilde{B}_*|^2\sum_{k=1}^{M}\Phi_k(1,s)-s\frac{\sigma^2}{\rho}|\widetilde{B}_*|^{\alpha}}\right]\frac{\dif s}{s}.
\end{align*}
By evaluating $\mathbb{E}_{|\widetilde{B}_*|^2}\left[\cdot\right]$ with $|\widetilde{B}_*|^2\sim\exp(\pi\lambda_{\Sigma})$ in the above integral, the result in \eqref{Eqn:DownlinkRate} is obtained. 

For the uplink rate $c_{ul}$, we can readily obtain its following expression based on the above results of the uplink rate:
\begin{align*}
c_{ul}=\sum_{m=1}^{M}\vartheta_m\int_{0}^{\infty}\left[1-\mathcal{L}_{G_*}\left(\frac{sQ}{w_m}\right)\right]\mathbb{E}\left[e^{-sI_{ul}\|\widetilde{B}_*\|^{\alpha}}\right]  \frac{\dif s}{s}.
\end{align*} 
Since $G_*\sim\text{Gamma}(N_m,\frac{1}{N_m})$, we know $\mathcal{L}_{G_*}(\frac{sQ}{w_m})=(1+\frac{sQ}{w_m})^{-N_m}$. Also, we have
\begin{align*}
&\mathbb{E}\left[e^{-sI_{ul}\|\widetilde{B}_*\|^{\alpha}}\right]=\mathbb{E}\left[e^{-sI_{ul}\|\widetilde{B}_*\|^{\alpha}}\big| \|\widetilde{B}_*\|^{2}=x\right]\\
&=\mathbb{E}\left[\exp\left(-s x^{\frac{\alpha}{2}}\sum_{U_j\in\mathcal{U}_s}\frac{QG_j}{\|\widetilde{B}_*-U_j\|^{\alpha}}\right)\right]\\
&\stackrel{(b)}{=}\mathbb{E}\left[\exp\left(-s x^{\frac{\alpha}{2}}Q\sum_{U_j\in\mathcal{U}_s} \frac{G_j}{(x+\|U_j\|^2)^{\frac{\alpha}{2}}}\right)\right]\\
&\stackrel{(c)}{\gtrapprox}e^{-\pi\lambda_{\Sigma} \sum_{m=1}^{M} \frac{q_m\vartheta_m}{w^{\frac{2}{\alpha}}_m} \int_{0}^{\infty }\left[\frac{sQ}{sQ+(1+\frac{r}{x})^{\frac{\alpha}{2}}}\right] \dif r}\\
&=\exp\left[-\pi\lambda_{\Sigma} x\sum_{m=1}^{M}w^{-\frac{2}{\alpha}}_m\Phi_m\left(1,\frac{sQw_m}{\eta(1-\rho)P_m}\right)\right],
\end{align*}
where $(b)$ is obtained based on the Slinvyka theorem and $(c)$ is obtained by assuming that all users in $\mathcal{U}_s$ form a homogeneous PPP of intensity $\sum_{m=1}^{M}q_m\lambda_{m}$. This yields $\mathbb{E}\left[e^{-sI_{ul}\|\widetilde{B}_*\|^{\alpha}}\right]\approx 1/[1+\sum_{m=1}^{M}w^{-\frac{2}{\alpha}}_m\Phi_m(1,\frac{sQw_m}{\eta(1-\rho)P_m})]$
 due to $|\widetilde{B}_*|^2\sim\exp(\pi\lambda_{\Sigma})$. Finally, \eqref{Eqn:UplinkRate} is acquired by substituting the results of $\mathcal{L}_{G_*}(sQ/w_m)$ and $\mathbb{E}\left[e^{-sI_{ul}\|\widetilde{B}_*\|^{\alpha}}\right]$ into the above expression of $c_{ul}$. 
  
\subsection{Proof of Theorem \ref{Thm:OptSol}}\label{App:ProofOptSol}
According to the optimization problem in \eqref{Eqn:OptProbEngEff}, there are two observations that can be drawn from the constraints in set $\mathcal{S}_{\rho,\beta}$: (i) Constraint $\mathbb{E}[E_{eh}]\geq (1-\beta)\tau Q$ can reduce to  $\frac{Q}{Q+\eta\mathbb{E}[P_{eh}]}\leq \beta\leq\overline{\beta}$ and $\rho\in\mathcal{S}'_{\rho}$ where $\mathcal{S}'_{\rho}\defn [\underline{\rho}, 1-\frac{(1-\beta)Q}{\eta\beta\mathbb{E}[P_{eh}]}]$. (ii) Constraint $\overline{\epsilon_{eh}}\geq \epsilon_{eh}$ can be expressed as $\overline{\epsilon_{eh}}\geq F_{P_{eh}}(\underline{P_{eh}})$ and $\mathcal{S}''_{\rho}\defn\{\rho\in(0,1): \overline{\epsilon_{eh}}\geq F_{P_{eh}}(\underline{P_{eh}})\}$ is a set with $\inf \mathcal{S}''_{\rho}=0$ and $\sup \mathcal{S}''_{\rho}<1$. Thereby, the feasible set of $\beta$ is $\mathcal{S}_{\beta}=[\frac{Q}{Q+\eta\mathbb{E}[P_{eh}]},\overline{\beta}]$ and the feasible set of $\rho$ is $\mathcal{S}_{\rho}\defn \mathcal{S}'_{\rho}\cap\mathcal{S}''_{\rho}$ that is always nonempty, which means we have a feasible set  $\mathcal{S}_{\rho,\beta}\equiv\mathcal{S}_{\rho}\times\mathcal{S}_{\beta}$. Next, the optimal solution pair ($\rho^{\star}$,$\beta^{\star}$) that exists in set $\mathcal{S}_{\rho,\beta}$ can be shown as follows. Consider the function $g$ of $x$ as $g(x)=\frac{ax+b}{c x+d}$ in which $a,b,c,d,x$ are all positive and real-valued. It then can be shown that $g(x)$ is a monotonic increasing (decreasing) function of $x$ if and only if $ad>bc$ ($ad<bc$). As a result, the objective function in \eqref{Eqn:OptProbEngEff} increases as $\beta$ increases if and only if $\frac{c_{dl}}{c_{ul}}>1+\frac{\sum_{m=1}^{M}\vartheta_m(P_m-Q)}{\sum_{m=1}^{M}\vartheta_mP_{m,on}+Q}$. In other words, if $\frac{c_{dl}}{c_{ul}}>1+\frac{\sum_{m=1}^{M}\vartheta_m(P_m-Q)}{\sum_{m=1}^{M}\vartheta_mP_{m,on}+Q}$ holds, increasing $\beta$ always increases the energy efficiency and this thus leads to the optimal value of $\beta$ that should happen at $\beta^{\star}=\overline{\beta}$. Also, since $c_{dl}$ depends on $\rho$ whereas $c_{ul}$ does not depend on it, increasing $\rho$ makes $\frac{c_{dl}}{c_{ul}}$ increase as well so as to improve the energy efficiency. Thus, if $ \overline{\mathcal{S}_{\rho}}\defn\{\rho\in\mathcal{S}_{\rho}:c_{dl}>c_{ul}(1+\frac{\sum_{m=1}^{M}\vartheta_m(P_m-Q)}{\sum_{m=1}^{M}\vartheta_mP_{m,on}+Q})\}$ is not empty, then the optimal value of $\rho$ must happen at $\rho^{\star}=\sup \overline{\mathcal{S}_{\rho}}$, and thereby we can conclude that $(\rho^{\star},\beta^{\star})=(\sup \overline{\mathcal{S}_{\rho}}, \overline{\beta})$ if $\overline{\mathcal{S}_{\rho}}$ is not empty. By the same reasoning, if $\underline{\mathcal{S}_{\rho}}\defn\{\rho\in\mathcal{S}_{\rho}:c_{dl}<c_{ul}(1+\frac{\sum_{m=1}^{M}\vartheta_m(P_m-Q)}{\sum_{m=1}^{M}\vartheta_mP_{m,on}+Q})\}$  is nonempty, then  
 $(\rho^{\star},\beta^{\star})=(\inf \overline{\mathcal{S}_{\rho}}, \frac{Q}{Q+\eta\mathbb{E}[P_{eh}]})$ if $\underline{\mathcal{S}_{\rho}}$ is nonempty. This completes the proof. 


\bibliographystyle{ieeetran}
\bibliography{IEEEabrv,Ref_HetNetsSWIPT}

\end{document}